\documentclass[]{emulateapj}
 \usepackage[colorlinks,citecolor=blue,linkcolor=blue]{hyperref}
\usepackage{etoolbox}

\makeatletter

\patchcmd{\NAT@citex}
{\@citea\NAT@hyper@{%
\NAT@nmfmt{\NAT@nm}%
\hyper@natlinkbreak{\NAT@aysep\NAT@spacechar}{\@citeb\@extra@b@citeb}%
\NAT@date}}
{\@citea\NAT@nmfmt{\NAT@nm}%
\NAT@aysep\NAT@spacechar\NAT@hyper@{\NAT@date}}{}{}

\patchcmd{\NAT@citex}
{\@citea\NAT@hyper@{%
\NAT@nmfmt{\NAT@nm}%
\hyper@natlinkbreak{\NAT@spacechar\NAT@@open\if*#1*\else#1\NAT@spacechar\fi}%
{\@citeb\@extra@b@citeb}%
\NAT@date}}
{\@citea\NAT@nmfmt{\NAT@nm}%
\NAT@spacechar\NAT@@open\if*#1*\else#1\NAT@spacechar\fi\NAT@hyper@{\NAT@date}}
{}{}

\makeatother
\usepackage{xspace}
\usepackage{apjfonts}
\usepackage{amssymb, amsmath}
\usepackage{natbibspacing, natbib}
\usepackage{aas_macros}

\newcommand{\myemail}{tleung@astro.cornell.edu}

\newcommand{\mbulge}{\mbox{$M_{\rm bulge}$}\xspace}

\newcommand{\Msun}{\mbox{$M_{\odot}$}\xspace}

\newcommand{\Lsun}{\mbox{$L_{\odot}$}\xspace}
\newcommand{\LIR}{\mbox{$L_{\rm IR}$}\xspace}
\newcommand{\LFIR}{\mbox{$L_{\rm FIR}$}\xspace}
\newcommand{\rarr}{$\rightarrow$}
\newcommand{\aco}{\mbox{CO($J$\,=\,1\,\rarr\,0)}\xspace}
\newcommand{\bco}{\mbox{CO($J$\,=\,2\,\rarr\,1)}\xspace}
\newcommand{\cco}{\mbox{CO($J$\,=\,3\,\rarr\,2)}\xspace}
\newcommand{\rot}[3][CO]{\mbox{#1($J$\,=\,#2\,\rarr\,#3)}}

\newcommand{\mgii}{\mbox{Mg{\scriptsize II}~2798\AA}\xspace}
\newcommand{\oiii}{\mbox{$[$O{\scriptsize III}$]$~4959, 5007\AA}\xspace}

\newcommand{\kms}{\mbox{km\,s$^{-1}$}\xspace}

\newcommand{\pmOne}{\mbox{$^{-1}$}\xspace}
\newcommand{\alphaco}{\mbox{$\alpha_{\rm CO}$}\xspace}
\newcommand{\alphaU}{\mbox{$M_{\odot}~($K\,\,km\,\,pc$^2)^{-1}$}\xspace}
\newcommand{\sfrU}{\mbox{\Msun\,yr$^{-1}$}\xspace}
\newcommand{\E}[1]{\mbox{$\times10^{#1}$}}
\newcommand{\petm}[2]{$^{+#1}_{-#2}$}
\newcommand{\eq}{\,=\,}
\newcommand{\ssim}{\,$\sim$\,}
\newcommand{\pmm}{\,$\pm$\,}
\newcommand{\eg}{{e.g.,~}}
\newcommand{\ie}{{i.e.,~}}
\newcommand{\Fig}[1]{Figure~\ref{fig:#1}}

\newcommand{\Tab}[1]{Table~\ref{tab:#1}}
\newcommand{\Sec}[1]{\S\ref{sec:#1}}
\newcommand\tna{\,\tablenotemark{a}}
\newcommand\tnb{\,\tablenotemark{b}}
\newcommand\tnc{\,\tablenotemark{c}}
\newcommand\tnd{\,\tablenotemark{d}}
\newcommand\tne{\,\tablenotemark{e}}

\def\spitzer {{\it Spitzer Space Telescope}\xspace}
\def\pdbi     {Plateau de Bure Interferometer\xspace}
\def\carma    {Combined Array for Research in Millimeter-wave Astronomy\xspace}

\newcommand{\ncode}[1]{{\sc #1}}
\newcommand{\uvmcmcfit}{\ncode{uvmcmcfit}\xspace}

\newcommand{\SF}{star formation\xspace}
\newcommand{\SB}{starburst\xspace}

\newcommand{\fir}{far-IR\xspace}

\newcommand{\mir}{mid-IR\xspace}

\slugcomment{Accepted to the ApJ}
\makeatletter
\renewcommand\normalsize{\@setfontsize\normalsize{10.56}{11.4}}
\makeatother

\shorttitle{Molecular Gas Kinematics of the Strongly-Lensed Quasar Host Galaxy RXJ1131$-$1231}
\shortauthors{Leung, Riechers \& Pavesi}

\begin{document}

\title{Molecular Gas Kinematics and Star Formation Properties of the Strongly-Lensed Quasar Host Galaxy RXS J1131$-$1231}
\author{T. K. Daisy Leung, Dominik A. Riechers, and Riccardo Pavesi}
\affil{Department of Astronomy, Space Sciences Building, Cornell University,
Ithaca, NY 14853, USA; \myemail}

\begin{abstract}
We report observations of \bco and \cco line emission towards
the quadruply-lensed quasar
RXS\,J1131$-$1231 at $z$\,=\,0.654 obtained using the
\pdbi (PdBI) and the \carma (CARMA).
Our lens modeling shows that the asymmetry in the double-horned \bco line profile
is mainly a result of differential lensing, where
the magnification factor varies from $\sim$3 to $\sim$9 across
different kinematic components.
The intrinsically symmetric line profile and a smooth source-plane velocity
gradient suggest that
the host galaxy is an extended rotating disk, with a CO size of $R_{\rm CO}$$\sim$6\,kpc and
a dynamical mass of $M_{\rm dyn}$\ssim8\E{10}\,\Msun.
We also find a secondary CO-emitting source near RXS\,J1131$-$1231 whose location is
consistent with the optically-faint companion reported in previous studies.
The lensing-corrected molecular gas masses are $M_{\rm gas}$\eq(1.4\pmm0.3)\E{10}\,\Msun
and (2.0\pmm0.1)\E{9}\,\Msun for RXS\,J1131$-$1231 and the companion, respectively.
We find a lensing-corrected stellar mass of
$M_{*}$\eq(3\pmm1)\E{10}\,\Msun and a star formation rate of SFR$_{\rm FIR}$\eq(120\pmm63)\,\sfrU,
corresponding to a specific SFR and star formation efficiency comparable to
$z$\ssim1 disk galaxies not hosting quasars.
The implied gas mass fraction of $\sim$18\pmm4\% is consistent with the previously-observed cosmic decline since $z$\ssim2. We thus find no evidence for quenching of star formation in RXS\,J1131$-$1231. This agrees with our finding of an elevated \mbox{$M_{\rm BH}$$/$$M_{\rm bulge}$} ratio of \mbox{$>$0.27\petm{0.11}{0.08}\%} compared to the
local value, suggesting that the bulk of its black hole
mass is largely in place while its stellar bulge is still assembling.
\end{abstract}


\section{Introduction}
Many recent studies of galaxy evolution have been focused on investigating the interplay between star formation and active galactic nucleus (AGN) activity across cosmic epochs \citep[e.g.,][]{DiMatteo05a,Alexander05a,Hopkins06b,Coppin08a,Page12a,Simpson12a,Lamastra13a}.
It is currently not well-understood when and how the supermassive black holes (SMBHs) and
stellar populations of present-day massive galaxies were assembled,
but it is clear that the co-moving \SF rate and the black hole accretion rate densities both
increased substantially since $z$\,$>\,3$ and reached their climax at $z$\ssim$2$, followed by
a rapid decline toward $z$\ssim$0$ \citep[\eg][]{Hopkins06a, Madau14a}.
A leading explanation for this decline is the decrease in molecular gas content and
\SF efficiency \citep[e.g.,][]{Erb06a, CW13, Walter14a},
but direct molecular gas measurements at intermediate redshift
$(0.2\,<\,z\,<\,1)$ that could confirm this explanation remain largely limited to
spatially unresolved CO observations of
a modest sample of $\sim$$30$ ultra-luminous infrared galaxies \citep[ULIRGs; ][]{Combes11a, Combes13a}.

Meanwhile, empirical scaling relations such as the
$M_{\rm BH}$$-$\mbulge relation \citep[e.g.,][]{Magorrian98a, HR04a}
have been established locally, suggesting a co-eval growth between local SMBHs and their host galaxies.
Attempts to extend this relation out to higher redshifts, beyond the peak epoch
of \SF and AGN activity, have been made in recent years.
These studies find that
high-$z$ AGN host galaxies do not appear to follow the same $M_{\rm BH}$$-$\mbulge relation as nearby spheroidal galaxies. \citep[\eg][]{Walter04a,Borys05a, McLure06a,Peng06a, Riechers08a,Coppin08a,Alexander08a}.
Yet, the $M_{\rm BH}$$-$\mbulge relation remains poorly-constrained
at intermediate redshifts due to the difficulty
in separating the stellar component contributing to the optical emission from that of the bright AGN.
This stems from the limited resolving power, which restricts the dynamic range that can be achieved at positions near the AGN.
Strong gravitational lensing provides the magnification necessary to spatially separate the AGN emission from the host galaxy stellar emission, significantly improving the dynamic range that can be achieved in studies of their host galaxies with current instruments.

\defcitealias{Sluse03a}{S03}
The quasar RXS J113151.62$-$123158 (hereafter RXJ1131)
at $z_\textrm{s, QSO}$\eq$0.658$ \citep[hereafter S03]{Sluse03a} is a particularly well-suited source for
studying the evolution of molecular gas properties in quasar host galaxies and the
connection between SMBHs and their host galaxies at intermediate redshift
given its unique lensing configuration.
The stellar emission in the host galaxy of RXJ1131 is lensed into
an Einstein ring of $1$\farcs$83$ in radius
that is clearly separated from the quadruply imaged quasar emission \citep[hereafter C06]{Claeskens06a}.
The foreground lens is an elliptical galaxy at $z_\textrm{L}$\eq$0.295$ \citepalias{Sluse03a}.
\citet[][]{Reis14a} report a high spin parameter of $a$\,$\sim$\,$0.9$ for the moderate-mass black hole in RXJ1131 \citep[$M_{\rm BH}$\eq$8$\E{7}\Msun;][]{Sluse12a},
with an intrinsic bolometric luminosity of $L_{\rm bol, X}$\eq$1.3$\E{45}\, ergs\,s\pmOne \citep{Pooley07a}.

In this paper, we present \bco and \cco line observations and
broadband photometry spanning rest-frame UV to radio wavelengths towards RXJ1131 to
study the properties of its molecular gas, dust and stellar populations.
In \Sec{obs}, we outline details of the observations and of our data reduction procedures.
In \Sec{results}, we report results for the \bco and \cco emission and broadband
photometry in RXJ1131.
In \Sec{anal}, we present lens modeling and dynamical modeling of the \bco data and
spectral energy distribution (SED) modeling of the photometric data.
In \Sec{diss}, we discuss
the ISM properties of the host galaxy of RXJ1131 and compare
them to other galaxy populations at low and high redshift.
Finally, we summarize the main results and present our conclusions in \Sec{sum}.
We use a concordance $\Lambda$CDM cosmology throughout this paper, with
parameters from the WMAP9 results:
$H_0$ = $69.32$ \kms Mpc\pmOne, $\Omega_{\rm M}$ = $0.29$, and
$\Omega_{\Lambda}$ = $0.71$ \citep{Hinshaw13a}.
\defcitealias{Claeskens06a}{C06}

\section{Observations} \label{sec:obs}

\subsection{CARMA \cco}
Observations of the \cco rotational line
($\nu_{\rm rest}$\,=\,345.79599\,GHz) towards RXJ1131 at $z_{\rm s, QSO}$\eq0.658
were carried out with the \carma (CARMA;
Program ID: cf0098; PI: D. Riechers) in the D array configuration.
The line frequency is redshifted to $\nu_{\rm obs}$\eq209.10443\,GHz at the quasar redshift.
Observations were carried out
on 2014 February 02 under poor 1.5\,mm
weather conditions and on 2014 February 17 under good 1.5\,mm
weather conditions.
This resulted in a
total on-source time of 2.94 hours after flagging poor
visibility data.
The correlator setup provides a bandwidth of 3.75 GHz in
each sideband and a
spectral resolution of 12.5 MHz ($\sim$17.9 \kms). The
line was placed in the lower sideband with the local oscillator tuned to $\nu_{\rm LO}\sim$216 GHz. The radio quasars J1127$-$189 (first track) and 3C273
(second track) were observed
every 15 minutes for pointing, amplitude, and phase calibration. Mars was
observed as the primary absolute flux calibrator and 3C279 was observed as
the bandpass calibrator for both tracks.

Given that the phase calibrator used for the first track was faint and was
observed under poor weather conditions and that the phase calibrator used for
the second track was far from our target source, the phase calibration is
subpar, with an rms scatter $\sim$50$\degr$ over a baseline length of $\sim$135\,m.
We thus conservatively estimate
a calibration accuracy of $\sim$40\% based on the flux scale uncertainties,
the gain variations over time, and the phase scatter on the calibrated data. We
therefore treat the derived \cco line intensity with caution and ensure that our physical interpretation
of this system and the conclusion of this paper do not rely on this quantity.

The \ncode{miriad} package was used to calibrate the visibility data.
The calibrated visibility data were
imaged and deconvolved using the CLEAN algorithm with ``natural'' weighting. This yields a synthesized clean
beam size of 3\farcs2 $\times$ 1\farcs9 at a position angle (PA) of 8\degr for the lower sideband
image cube. The final rms noise is $\sigma$ = 13.3 mJy beam$^{-1}$
over a channel width of 25\,MHz. An rms noise of
$\sigma$\,=\,0.83\,mJy\,beam\pmOne is reached by averaging over the
line-free channels in both sidebands.

\subsection{PdBI \bco}
Observations of the \bco rotational line
($\nu_{\rm rest}$\,=\,$230.53800$ GHz)
towards RXJ1131
were carried out using the IRAM \pdbi (PdBI; Program ID: S14BX; PI: D.
Riechers).
Based on the CARMA \cco line redshift of $z_{\rm CO(3-2)}$\eq0.655,
the \bco line is redshifted to $\nu_{\rm obs}$\eq139.256\,GHz.
Two observing runs were carried out on 2014 December 06 and 2015
February 05 under good weather conditions in the C and D array configurations,
respectively.
This resulted in 3.75 hours of cumulative six antenna-equivalent on-source
time after discarding unusable visibility data.
The 2 mm receivers were used to cover the redshifted \bco line
and the underlying continuum emission, employing a correlator setup that provides
an effective bandwidth of 3.6 GHz (dual polarization) and a native spectral resolution of 1.95\,MHz
($\sim$\,$4.2$\,\kms).
The nearby quasars B1127$-$145 and B1124$-$186 were observed every 22 minutes
for pointing, secondary amplitude, and phase calibration, and B1055$+$018 was
observed as the bandpass calibrator for both tracks.
MWC349 and 3C279 were observed as primary flux calibrators for the C and D
array observations, respectively, yielding calibration accuracy better than 15\%.

The \ncode{gildas} package was used to calibrate and analyze the visibility data.
The calibrated visibility data were imaged and deconvolved using the CLEAN algorithm with ``natural''
weighting. This yields a synthesized clean beam size of 4$\farcs$44 $\times$ 1\farcs95 (PA = 13\degr).
The final rms noise is $\sigma$\,=\,1.45\,mJy\,beam\pmOne
over 10 MHz (21.5\,\kms). The continuum image at $\nu_{\rm cont}\sim$139\,GHz
is produced by averaging over 3.16\,GHz of line-free bandwidth. This
yields an rms noise of 0.082 mJy\,beam$^{-1}$.

\subsection{Karl G. Jansky Very Large Array (Archival)}
Our analysis also uses archival data of the 4.885\,GHz
radio continuum obtained with the
Karl G. Jansky Very Large Array (VLA; Program ID: AW741; PI: Wucknitz).
Observations were carried out on 2008 December 29 under excellent weather
conditions in the A array configuration for a total of $\sim$7 hours on-source time. The C-band receivers were used with a continuum mode setup,
providing a bandwidth of 50 MHz for the two IF bands with full polarization.
The nearby radio quasar J1130$-$149 was observed every 10 minutes for
pointing, amplitude, and phase calibration. J1331$+$305 was observed as the
primary flux calibrator, and J0319$+$415 was observed as the bandpass
calibrator, yielding $\sim$10\% calibration accuracy.
We used \ncode{aips} to calibrate the visibility data.
The calibrated visibility data were imaged and deconvolved using
the CLEAN algorithm with robust\,=\,0, which
was chosen to obtain a higher resolution image given high SNR.
This yields a synthesized clean
beam size of 0$\farcs$49 $\times$ 0\farcs35 (PA\,=\,0.18$\degr$) and a final
rms noise of $\sigma$ = 13 $\mu$Jy beam\pmOne.

\subsection{HST (Archival)}
We obtained an {\it HST} image taken with
the ACS
using the F555W filter ($V$-band)
from the
Hubble Legacy Archive.
The details of the observations can be found
in \citetalias{Claeskens06a}.
We apply an astrometric correction to the optical image by adopting the VLA 5\,GHz map as the
reference coordinate frame.
We shift the latter to the east by 0\farcs5963 in R.A. and $+$0\farcs8372 in
Dec., which is consistent with the typical astrometric precision (1$^{\prime\prime}-$2$^{\prime\prime}$) of
images from the Hubble Legacy Archive\footnote{http://hla.stsci.edu/hla\_faq.
html}. This astrometric correction is critical to avoid artificial spatial
offsets between different emitting regions and to carry out our lens modeling,
in which the absolute position of the foreground lensing galaxy is based on
its coordinates in the high-resolution optical image.
The VLA image is calibrated using a well-monitored phase
calibrator, with absolute positional accuracy of $\sim$2 mas.
For this reason, the absolute alignment between the VLA image and other
interferometric images reported in this paper are expected to have an astrometric
precision better than 0\farcs1, modulo uncertainties related to the SNR and phase
instability.

\section{Results} \label{sec:results}
\subsection{\bco Emission} \label{sec:CO21}
\begin{figure}[htbp]
\centering
\includegraphics[width=0.45\textwidth]{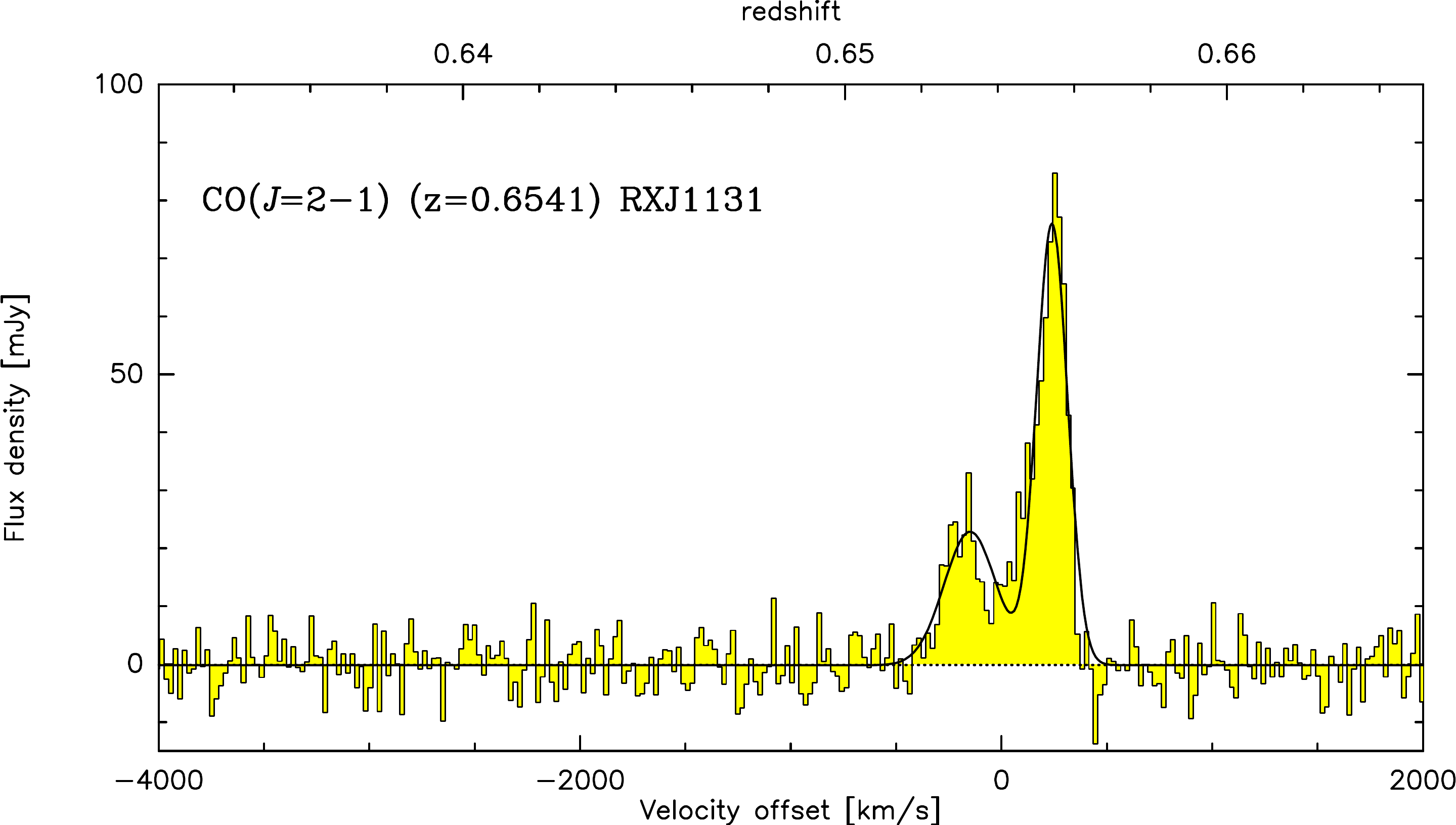}
\caption{Continuum-subtracted spectrum (histogram) of \bco emission towards RXJ1131, with a spectral resolution of
22\,\kms. The solid black line shows a double-Gaussian fit to the line profile.
The velocity scale is with respect to $z$\,=\,0.6541, which corresponds to the line center of RXJ1131 based on
the de-lensed line profile (\Fig{delensed}).
A detailed discussion of this effect is presented in \Sec{differential}. \label{fig:CO21spec}}
\end{figure}

We detect \bco line emission towards the background source RXJ1131 in the PdBI data
at $\gtrsim$\,27$\sigma$ significance. Based on this measurement, we refine the redshift of RXJ1131 to
$z_{\rm CO}$\eq0.6541\,$\pm$\,0.0002\footnote{This redshift is derived by fitting a double-Gaussian to
the de-lensed spectrum (\Fig{delensed})
instead of the observed spectrum (\Fig{CO21spec}) to avoid biases in our
redshift determination
due to differential lensing (see \Sec{differential}).
}. The emission is spatially and kinematically resolved
with a highly asymmetric double-horned line profile
as shown in \Fig{CO21spec}. Fitting a double Gaussian results in peak
flux densities of 75.3\pmm2.6 mJy and 24.0\pmm2.0 mJy, and a FWHM of
179\pmm9 \kms and 255\pmm28 \kms for the two components, respectively. The peaks are separated by
$\Delta v_{\rm sep}$\eq400\pmm12\,\kms. The total integrated line flux is 20.3\,$\pm$\,0.6 Jy \kms.

\begin{deluxetable}{lcc}[!htbp]
\tabletypesize{\scriptsize}
\tablecolumns{3}
\tablecaption{Observed Properties of RXJ1131 and its companion}
\tablehead{
\colhead{Parameter} &
\colhead{Unit} &
\colhead{Value}
}
\startdata
$z_{\rm CO(2-1)}$   &                          & 0.6541\pmm0.0002 \\ [0.5ex]
$I_{\rm CO(2-1)}$     & Jy\,km s\pmOne           & 20.3\pmm0.6 \\ [0.5ex]
$S_{\rm CO(2-1)}$\tna     & Jy km s\pmOne beam\pmOne & 8.12\pmm0.30 \\ [0.5ex]
FWHM$_{\rm CO(2-1)}$ \tnb & km s\pmOne               & 179\pmm9, 255\pmm28 \\ [0.5ex]
FWHM$_{\rm CO(2-1)}$\tnc        & km s\pmOne               & 220\pmm72 \\ [0.5ex]
$I_{\rm CO(3-2)}$     & Jy\,km s\pmOne           & 35.7\pmm6.9
\enddata
\label{tab:obsprop}
\tablenotetext{a}{Peak flux density in the intensity map.}
\tablenotetext{b}{From fitting a double Gaussian to the observed \bco spectrum (\Fig{CO21spec}).}
\tablenotetext{c}{From fitting a double Gaussian with a common FWHM to the de-lensed \bco spectrum (\Fig{delensed}).}
\end{deluxetable}

\begin{figure*}[!htbp]
\hspace{0.5em}
\includegraphics[trim=0 10 15 0, clip, width=0.452\textwidth]{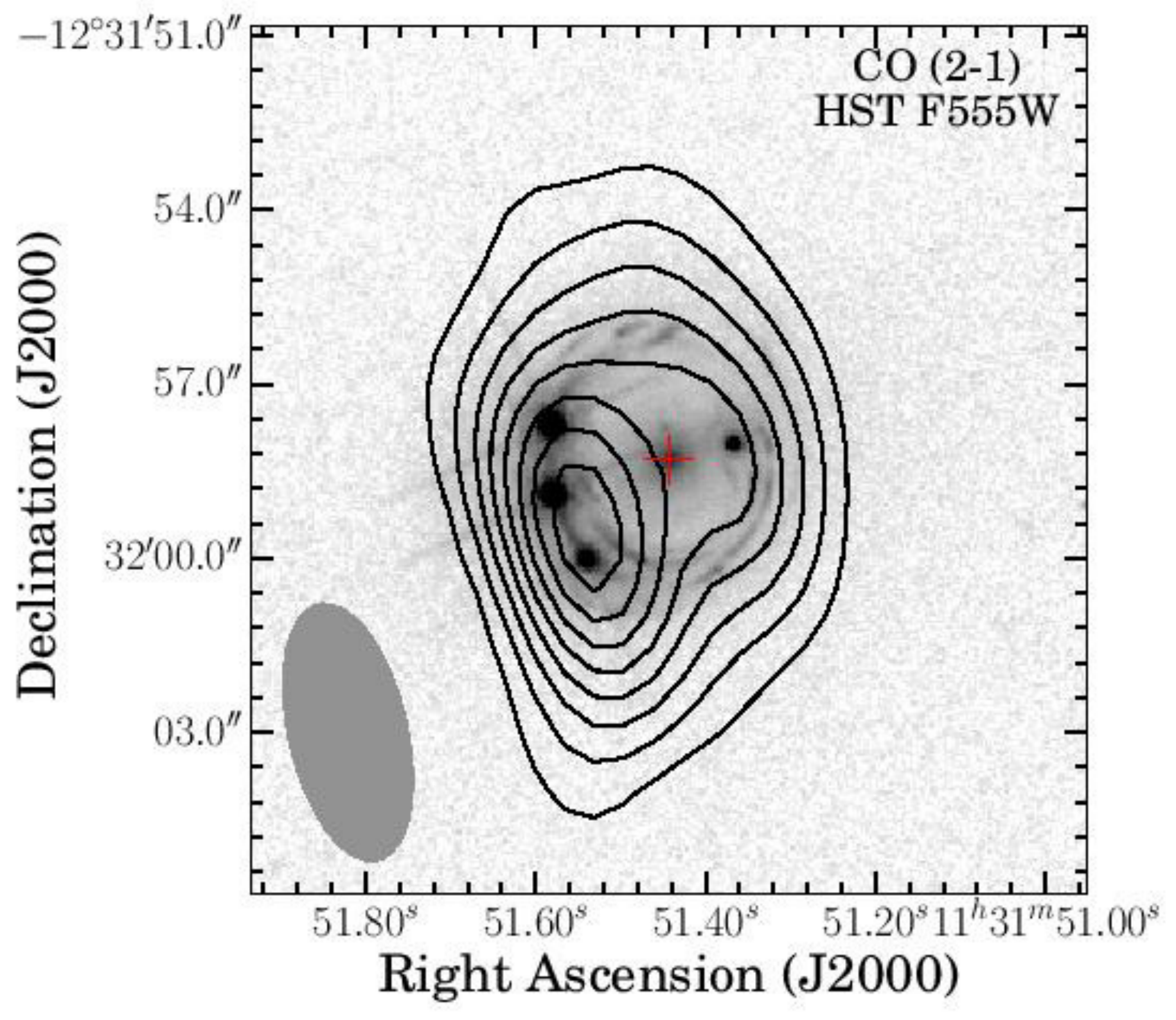}
\includegraphics[trim=5 -18 0 0, clip, width=0.462\textwidth]{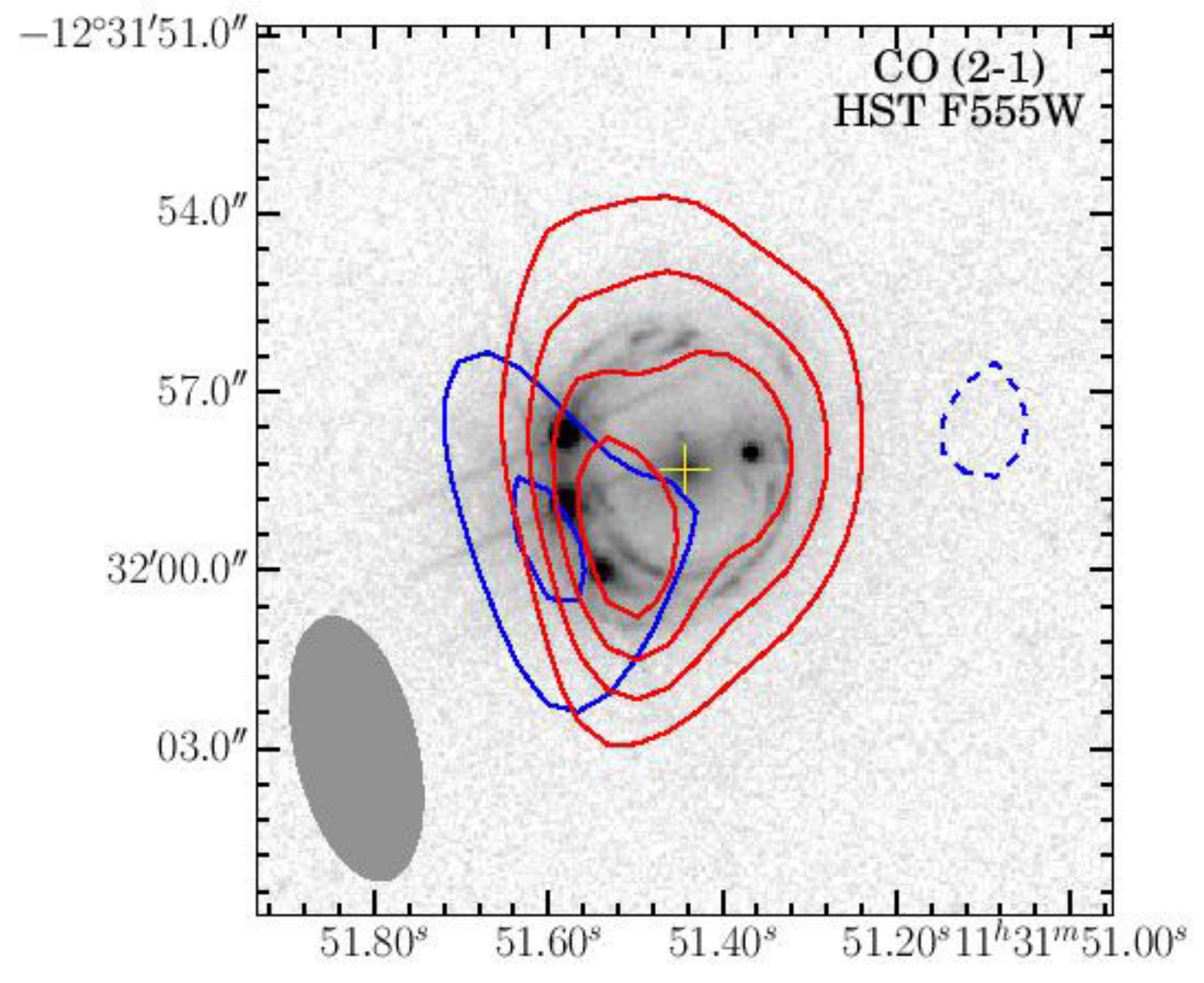}
\\
\includegraphics[trim=25 16 0 10, clip, width=0.85\textwidth]{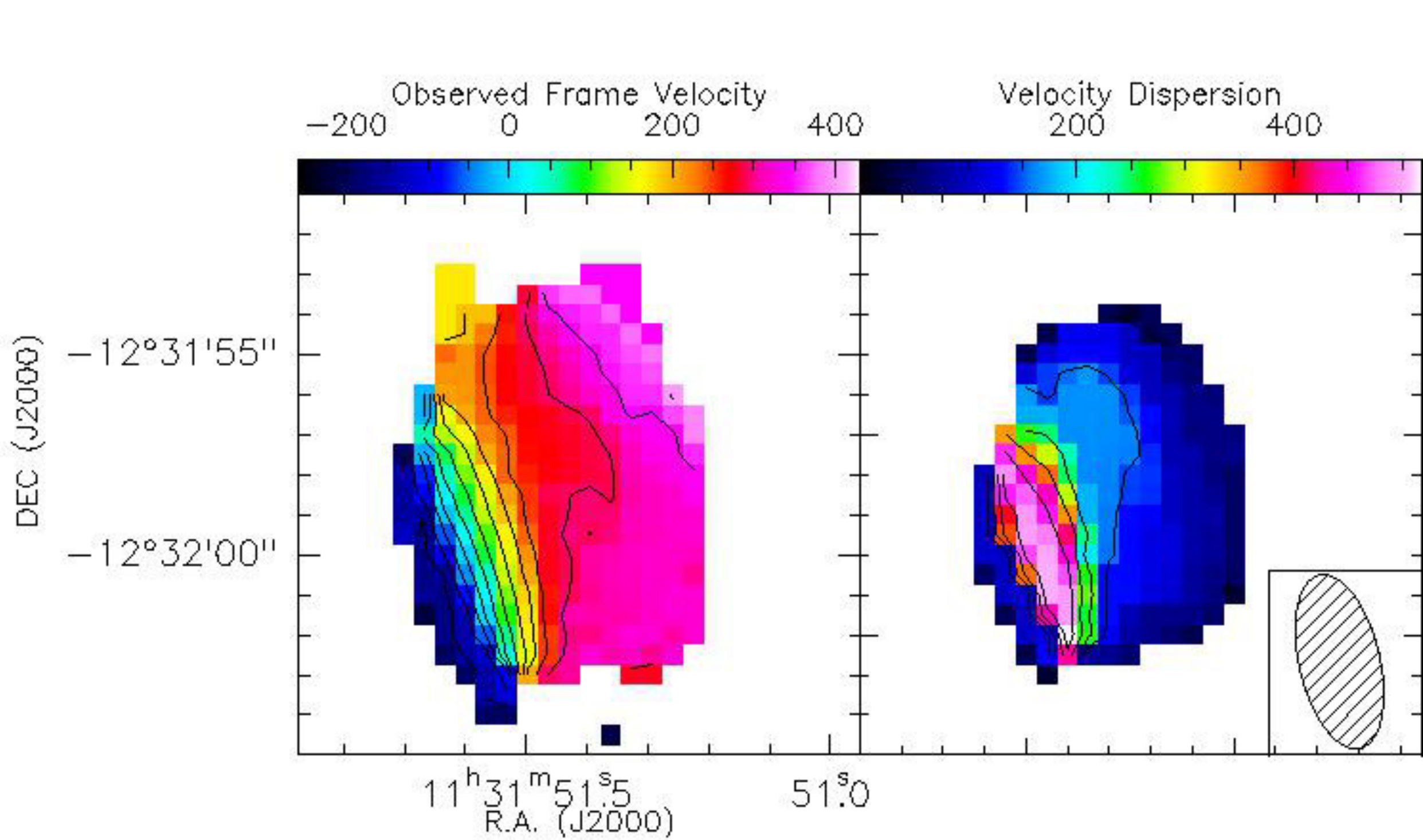}
\vspace{0.1em}
\caption{Top left: overlay of the velocity-integrated \bco emission on an archival {\it HST} $V$-band (F555W)
image.
Top right: same as top left, except the contours are color-coded to represent the red- and blueshifted emission, which
are extracted by integrating over $v$\,$\in$[$-$19, 357] \kms and $\in$[$-$395, $-$19] \kms, respectively.
The contours in both top panels start at 3$\sigma$ and increment in steps of
$\pm$3$\sigma$, where $\sigma$\,=\,0.3 mJy beam\pmOne for the top left panel,
and
$\sigma$ = 0.4 mJy beam\pmOne (red) and 0.5 mJy beam\pmOne (blue)
for the top right panel.
The crosses denote the
location of the foreground galaxy at $z$\,=\,0.295.
Contours for the first (bottom left) and second (bottom right) moment maps of the \bco line emission
are shown in steps of
50 \kms, and 100 \kms, respectively.
The synthesis beam size is 4\farcs4 $\times$ 2\farcs0, at PA = 13$\degr$.
\label{fig:CO21mom}}
\end{figure*}

We construct a zeroth order moment map, red/blue channel maps, and
first and second moment maps, as shown in \Fig{CO21mom},
using the $uv$-continuum subtracted data cube over a velocity range of
$\Delta v$ $\sim$ 750\,\kms.
The higher-order moment maps are produced using
unbinned channel maps with 3$\sigma$ clipping.
The peak flux density is 8.12\pmm0.30 Jy\,\kms\,beam\pmOne
in the intensity map.
Observed properties of the \bco emission line are summarized in \Tab{obsprop}.

The deconvolved source size FWHM obtained from fitting a single two-dimensional Gaussian
to the integrated line emission in the image plane yields 5\farcs1\pmm0\farcs7$\times$3\farcs7\pmm0\farcs7,
which is consistent with that obtained by visibility-plane fitting within the uncertainties.
Since the spatial distribution of the observed CO emission is unlikely to be fully described by a simple Gaussian
and appears to be a superposition of at least two components (top left panel of \Fig{CO21mom}),
we also fit two Gaussians to the intensity map.
This yields deconvolved source sizes of
3\farcs8\pmm0\farcs4$\times$1\farcs9\pmm0\farcs4 and 3\farcs6\pmm0\farcs3$\times$1\farcs5\pmm0\farcs3, separated by $\sim$2\farcs2 in RA and $\sim$1\farcs7 in Dec.
The deconvolved source sizes of both models suggest that the gravitationally lensed CO emission is more extended
than the optical
``Einstein ring'', which has a diameter of $\sim$3\farcs6
(i.e., the ``Einstein ring'' formed by CO emission is likely to have a larger diameter compared to the optical one).
This is consistent with the centroid position of the redshifted emission, which is along the quasar arc seen in the optical image,
and the blueshifted emission, which is offset further to the SE (top right of \Fig{CO21mom}).
Therefore, the CO-emitting region in RXJ1131 is likely to be more extended than its stellar and quasar emission.

We also place an upper limit on \rot[HNC]{2}{1} line emission
in the foreground galaxy at $z\sim$0.295.
Assuming a typical line width of 300\,\kms, this corresponds to a 3$\sigma$
limit of 0.35\,Jy\,\kms\,beam\pmOne.

\subsection{\cco Emission}
\begin{figure}[!htbp]
\includegraphics[width=0.455\textwidth]{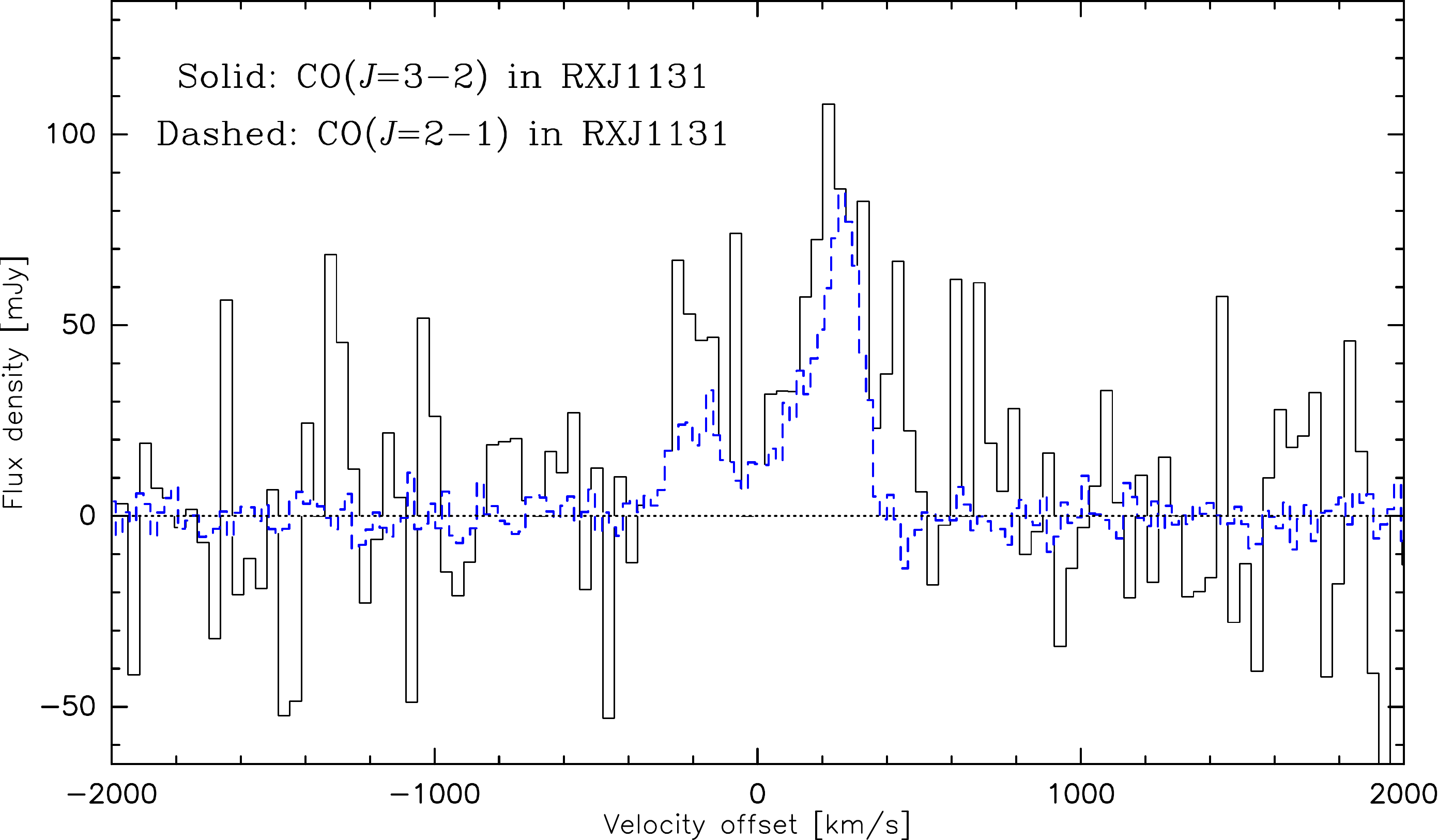}
\caption{CARMA \cco line profile (solid) without continuum subtraction is
over-plotted on the continuum-subtracted PdBI \bco line profile (dashed).
The velocity scale is the same as in \Fig{CO21spec}.
The spectral resolution for \cco and \bco
is 36 \kms and 22 \kms, respectively.
\label{fig:co32spec}}
\end{figure}

We detect \cco line emission towards RXJ1131 in the CARMA data at $\gtrsim$\,5$\sigma$ significance.
The \cco spectrum appears to be consistent with a double-peaked profile, as shown in \Fig{co32spec}, where
we over-plot spectra of the \bco and \cco lines.
We extract the peak fluxes and their corresponding uncertainties for the blue and red wing independently.
We find a peak line flux of 5.13\pmm1.43 Jy\,\kms beam\pmOne for the blue wing, indicating a $\gtrsim$3$\sigma$ detection for this component alone, and a peak line flux of 11.45\pmm1.99\,Jy\,\kms beam\pmOne for the red wing,
indicating a $\sim$\,6$\sigma$ detection.
We measure a line intensity of 35.7\,$\pm$\,6.9\,Jy\,\kms (\Tab{obsprop}) by summing up fluxes over the FWZI
linewidth used to infer the \bco line intensity.

Assuming that the spatial extent of \bco and \cco is similar and therefore the emission is
magnified by the same amount, the measured line intensities
correspond to a brightness temperature ratio of
$r_{\rm 32}$\,=\,$T_{\small \cco}$$/$$T_{\small \bco}$\,=\,0.78\,$\pm$\,0.37.
The quoted error bar is derived by adding the uncertainties associated with the CO line intensities
and those from absolute flux calibrations in quadrature.
This brightness temperature ratio is consistent with
thermalized excitation within the uncertainties, as commonly observed in nuclear regions of
nearby ULIRGs and high-$z$ quasars \citep[e.g.,][]{Weiss07a, Riechers11b, CW13}, but also
with the lower excitation seen in normal star-forming disks \citep[e.g.,][]{Dannerbauer09a, CW13, Daddi15a}.

\subsection{Continuum Emission} \label{sec:deblend}
No 1.4\,mm continuum emission is detected at the position of \cco
down to a 3$\sigma$ limit of 2.49\,mJy beam\pmOne.
This is consistent with the spectrum shown in \Fig{co32spec}.

\begin{figure}[!htbp]
\includegraphics[trim=12 27 0 1, clip, width=0.47\textwidth]{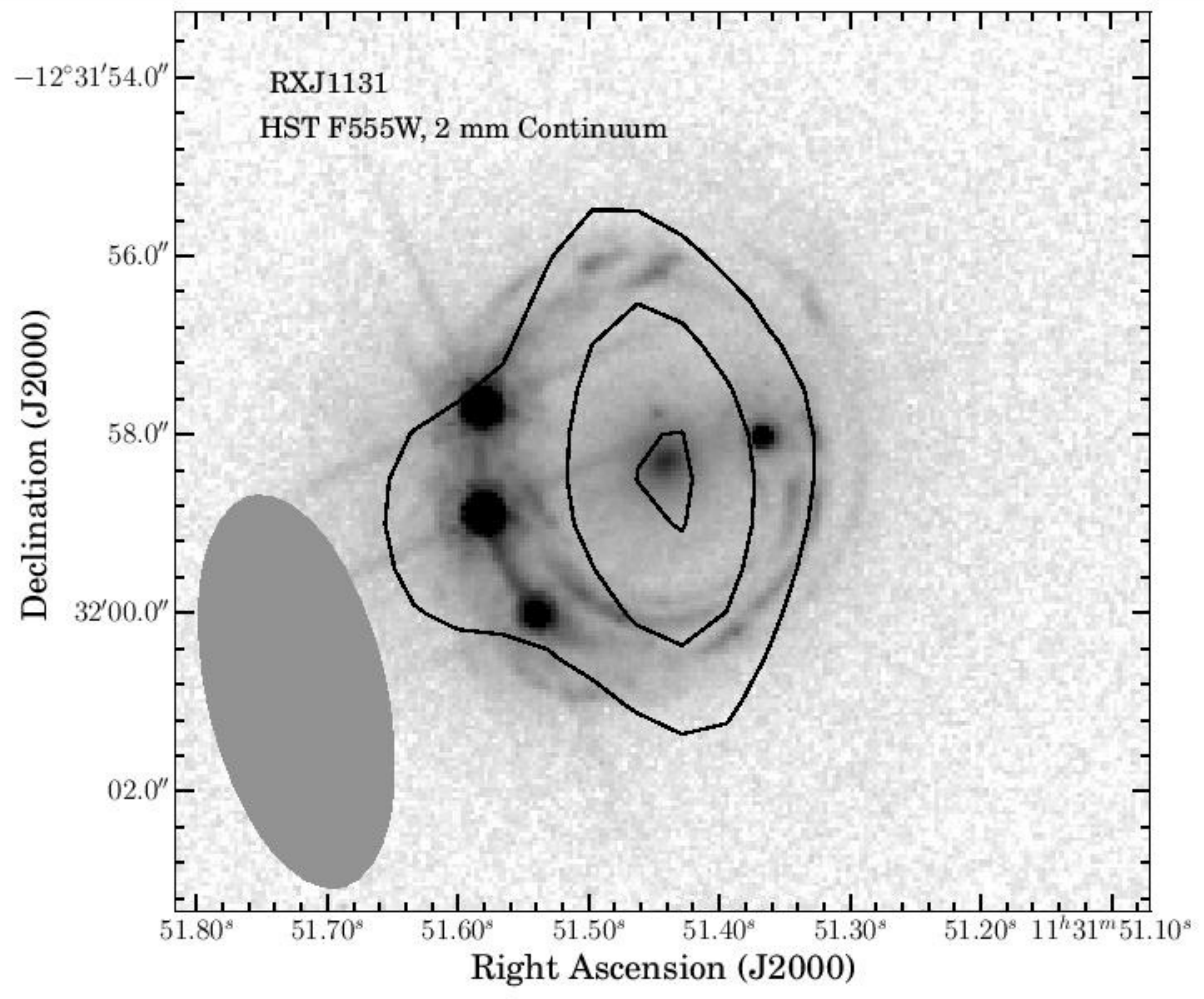}
\includegraphics[trim=10 5 0 0, clip, width=0.47\textwidth]{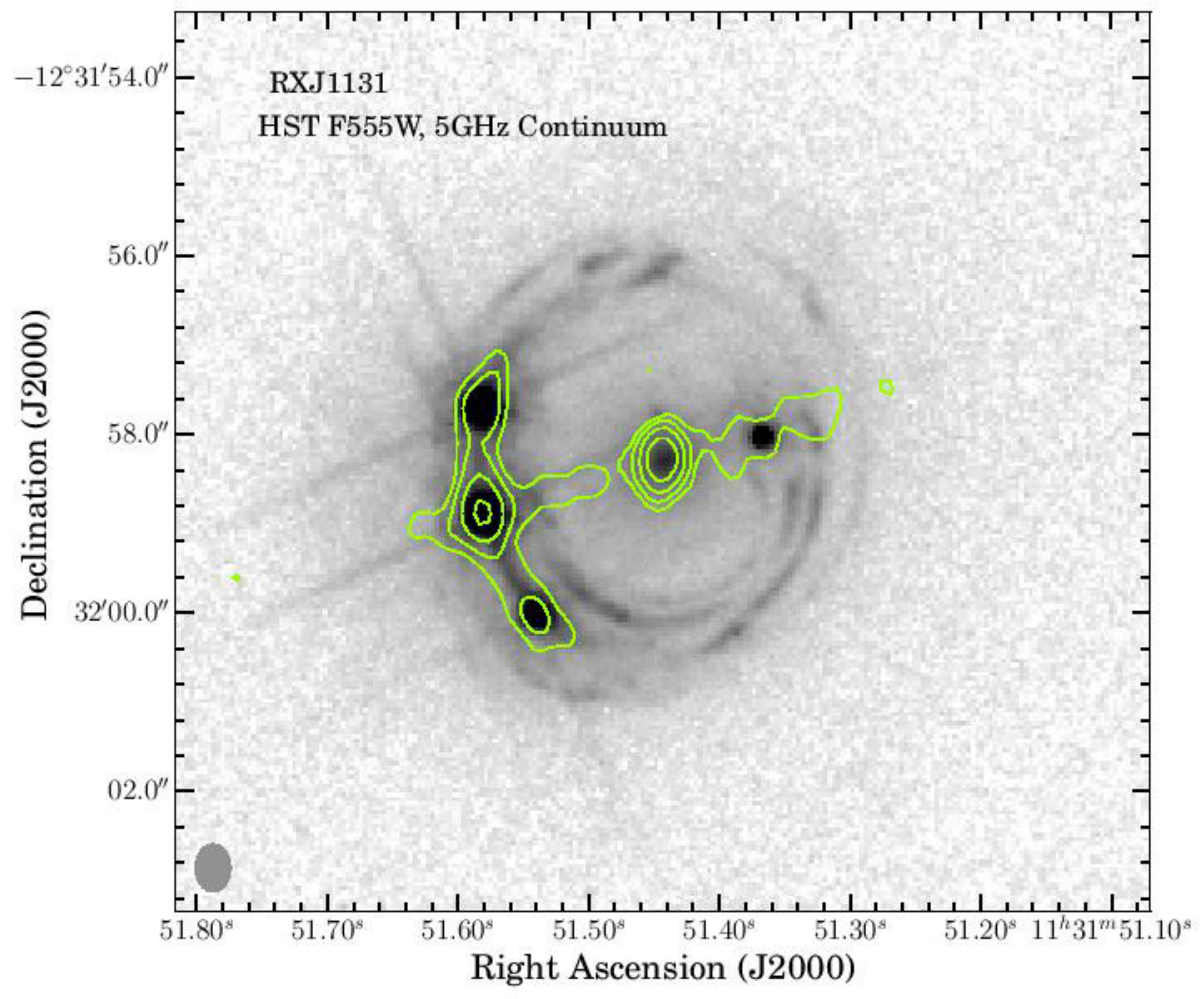}
\caption{Top: overlay of the PdBI 2\,mm continuum emission on the optical image.
Contours start and increment in steps of
$\pm$3$\sigma$, where $\sigma_{\rm 2mm}$\,=\,0.082 mJy beam\pmOne.
Bottom: overlay of the VLA 5\,GHz continuum emission on the optical image.
Contours correspond to $\pm2^n\sigma$, where
$\sigma_{\rm 5GHz}$\,=\,13 $\mu$Jy beam\pmOne
and $n$ is an integer running from 2 to 5.
Radio emission towards the foreground radio core is detected at $\gtrsim$57$\sigma$ significance.
The synthesis beam size is 4\farcs4 $\times$ 2\farcs0, at PA = 13$\degr$ for
the PdBI observations (top), and
0\farcs5 $\times$ 0\farcs4 (PA = 0.18$\degr$) for
the VLA observations (bottom).
\label{fig:cont}}\vspace{0.51em}
\end{figure}

We detect 2.2\,mm continuum emission at an
integrated flux density of
1.2\pmm0.2 mJy, with a peak flux of
$S_{\nu}$\,=\,799\pmm88\,$\mu$Jy\,beam\pmOne
centered on the lensing galaxy (\Fig{cont}).
Slightly extended emission
along the lensing arc is also detected.
This suggests that we detect emission in both
the foreground and the background galaxy and that the
emission is marginally resolved along its major axis.
We subtract a point source model in the visibility-plane to remove the unresolved part of the
emission, which we here assume to be dominated by the foreground galaxy. The emission
in the residual map coincides spatially with the lensing arc. We measure a flux density
of $S_{\nu}$\eq0.39\,$\pm$\,0.08\,mJy for this residual component.
This flux density is consistent with
the difference between the integrated and the peak flux density measured in the
original continuum map ($\sim$0.4 mJy).
We therefore adopt $S_{\nu}$\eq0.39\pmm0.12 mJy as the best estimate for the 2\,mm continuum flux of
the background galaxy (RXJ1131).
We here quote a conservative error bar, which is derived by adding the uncertainty
associated with the flux density of the
point-source model ($\delta S_{\nu}$\eq0.088\,mJy) with
that of the peak flux in the residual map (0.08\,mJy)
in quadrature. We caution that this does not account for the systematic uncertainties of the
de-blending procedure, where we have assigned 100\% of the point source flux to the foreground galaxy.
We report the peak flux in the original map
($S_{\nu}$\,=\,799\pmm88\,$\mu$Jy\,beam\pmOne) for the foreground galaxy, which
is the best estimate possible at the resolution of our observations, but we acknowledge that a non-negligible contribution from the background source to the peak flux cannot be ruled out.

The VLA C-band continuum image in \Fig{cont} shows resolved emission from the
jets and the core of the foreground elliptical galaxy
as well as emission toward the background quasar.
Multiple peaks are seen along the arc with their centroids
coincident with the optical emission from the quasar.
We extract the flux densities for the lensing arc and the radio core in \Tab{photometry}.
We find a spectral index of $\alpha^{\rm 2mm}_{\rm 6cm}$\,=\,$-$0.02\pmm0.07
for the foreground galaxy and $\alpha^{\rm 2mm}_{\rm 6cm}$\,=\,$-$0.35\pmm0.21
for the background galaxy by fitting
power-laws ($S_\nu \propto \nu^{\alpha}$) to their continuum fluxes at
5\,GHz and 2\,mm.
The spectral slope derived for the background source is flatter than the typical slope of pure synchrotron emission \citep[$\alpha$\ssim$-$0.7; e.g.,][]{Andreani93a}. This likely suggests
that at least a fraction of the observed 2\,mm emission arises from thermal dust emission.
This spectral slope would be even shallower
if the background source contributes to the unresolved fraction of the
2\,mm flux.
In this case, the 2\,mm flux of the foreground galaxy would be lower than the value reported here and
lead to a slope steeper than $\alpha^{\rm 2mm}_{\rm 6cm}$\ssim$-$0.02, which is flatter than
that typical of elliptical galaxies.
Assuming a spectral slope of $\alpha$\ssim$-$0.7 to account for synchrotron radiation in RXJ1131, we expect
a flux density of $S_{\rm 2mm}$\eq0.122\pmm0.004 mJy at 2\,mm.
The flux excess of $S_{\rm 2mm}$\eq0.27\pmm0.08\,mJy therefore likely arises due to thermal dust emission.

\subsection{Photometry} \label{sec:photometry}
We compile \mir (MIR) to \fir broadband photometry from various
catalogs available on the NASA/IPAC Infrared Science
Archive (IRSA) in \Tab{photometry} with aperture corrections
when warranted. These data were obtained using
the Cerro Tololo Inter-American Observatory (CTIO) for the Two Micron All Sky Survey \citep[2MASS;][]{Skrutskie06a},
the Wide-field Infrared Survey Explorer \citep[{\it WISE};][]{Wright10a},
the {\it Infrared Astronomical Satellite}
\citep[{\it IRAS};][]{Neugebauer84a}, and
the Multiband Imaging Photometer \citep[MIPS;][]{Rieke04a} and
Mid-infrared Infrared Array Camera \citep[IRAC;][]{Fazio04a} on
the \spitzer.
We retrieve PBCD (level 2) {\it Spitzer}/IRAC images from the
Spitzer Heritage Archive and perform aperture photometry on
the channel 1 image to extract the flux density at 3.6\,$\mu$m
since it is not available from the IRSA archive.

The emission in the IRAC images is slightly extended. We thus use an
{\it HST} image ($\sim$0\farcs07 resolution) to determine the
origin of their centroids, all of which are found to be
centered at the position corresponding to the lensed emission from the
background galaxy. To recover the diffuse background emission, we subtract a
point source model centered on the lensing galaxy, using the average
FWHM found by fitting a Gaussian profile to several field stars
with the \ncode{imexam} routine of IRAF.
We perform aperture photometry on the residual image
to obtain decomposed flux measurements of the background galaxy.
The photometry for the foreground galaxy is then obtained
by subtracting the background emission from the
observed total flux. The resulting photometry in
\Tab{photometry} is obtained after performing an aperture correction
described in the IRAC Instrument Handbook\footnote{http://irsa.ipac.caltech.edu/data/SPITZER/docs/irac/iracinstrumenthandbook/} to
correct for the fact that the imaging was calibrated
using a 12$^{\prime\prime}$ aperture, which is larger than the aperture (5\farcs8) we used to
perform aperture photometry.

We fit a power-law spectrum to the
decomposed IRAC photometry to disentangle the background and foreground
emission from the total flux observed in the MIPS 24\,$\micron$ band.
The spectral indices corresponding to the best-fitting curves are $\alpha$\,=\,$-1.8$ and
$\alpha$\,=\,$-0.85$ for the lensing galaxy and RXJ1131, respectively.
The latter
is consistent with the mean 3.6\,$-$\,8\,$\micron$
spectral slope of
$\alpha$\,=\,$-$1.07\,$\pm$\,0.53 found for unobscured AGN
\citep{Stern05a}. An extrapolation of the fit to 24\,$\micron$
yields 33.96\,$\pm$\,0.01\,mJy and 25.19\,$\pm$\,0.03\,mJy
for the foreground galaxy and RXJ1131, respectively.
The uncertainties are the standard deviations of
the extrapolated fluxes obtained from two independent Monte Carlo
simulations, each of 500 iterations.
We incorporate the decomposed 24\,$\micron$ data in our
SED fitting to provide some constraints on
the Wien tail beyond the dust peak
of the SED of RXJ1131.
Details of the SED modeling are presented in \Sec{SED}.

Extraction of the {\it Herschel}/SPIRE photometry at 250, 350, and 500\,$\micron$ was
carried out using \ncode{sussextractor} within the Herschel Interactive
Processing Environment \citep[HIPE;][]{Ott10a}
on Level 2 maps obtained from the Herschel Science Archive.
These maps were processed by the SPIRE pipeline
version 13.0 within HIPE. The \ncode{sussextractor} task estimates
the flux density from an image convolved with a kernel
derived from the SPIRE beam. The flux densities
measured by \ncode{sussextractor} were confirmed by
using the Timeline Fitter, which performs photometry
by fitting a 2D elliptical Gaussian to the Level 1 data at the
source position given by the output of \ncode{sussextractor}. The fluxes
obtained from both methods are consistent within the uncertainties.

\begin{deluxetable}{lccc}[!htbp]
\centering
\tabletypesize{\scriptsize}
\tablecolumns{4}
\tablecaption{Photometry data}
\tablehead{\colhead{Wavelength} & \colhead{Frequency} & \colhead{Flux Density} & \colhead{Instrument}\\
\colhead{($\micron$)} & \colhead{(GHz)} & \colhead{(mJy)} & \colhead{ } \vspace{0.05in}
\\  \cline{1-4} \vspace{-0.05in} \\
\multicolumn{4}{c}{Combined/Unresolved}
}
\startdata
1.25    & 239834  & 1.009 $\pm$ 0.090    & CTIO/J-Band \\
1.65    & 181692  & 1.448 $\pm$ 0.121    & CTIO/H-Band \\
2.17    & 138153  & 2.064 $\pm$ 0.160    & CTIO/Ks-Band \\
3.4     & 88174.2 & 7.027 $\pm$ 0.142    & {\it WISE}/W1 \\
3.6     & 83275.7 & 5.618 $\pm$ 0.002   & {\em Spitzer}/IRAC \\
4.5     & 66620.5 & 7.803 $\pm$ 0.002   & {\em Spitzer}/IRAC \\
4.6     & 65172.3 & 8.872 $\pm$ 0.163   & {\it WISE}/W2 \\
5.8     & 51688.4 & 10.720 $\pm$ 0.005  & {\it Spitzer}/IRAC \\
8.0     & 37474.1 & 14.470 $\pm$ 0.004  & {\it Spitzer}/IRAC \\
12      & 24982.7 & 21.960 $\pm$ 0.425  & {\it WISE}/W3 \\
12      & 24982.7 & $<$400              & {\it IRAS} \\
22      & 13626.9 & 55.110 $\pm$ 1.878  & {\it WISE}/W4 \\
24      & 12491.4 & 70.204 $\pm$ 0.026  & {\it Spitzer}/MIPS \\
25      & 11991.7 & $<$ 500             & {\it IRAS} \\
60      & 4996.54 & $<$ 600             & {\it IRAS} \\
100     & 2997.92 & $<$ 1000            & {\it IRAS} \\
250     & 1199.17 & 289.4 $\pm$ 9.6     & {\it Herschel}/SPIRE \\
350     & 856.55  & 168.2 $\pm$ 8.6     & {\it Herschel}/SPIRE \\
500     & 599.585 & 56.8 $\pm$ 8.8      & {\it Herschel}/SPIRE \\
1387.93 & 216     & $<$2.49             & CARMA \\
2152.82 & 139.256 & 1.23 $\pm$ 0.22   & PdBI \\
\cutinhead{Foreground Lensing Galaxy (deblended bands)} \\ [-1.5ex]
0.555   & 540167  & 0.056 $\pm$ 0.006   & {\it HST}-ACS/V-Band \\
0.814   & 368295  & 0.238 $\pm$ 0.013   & {\it HST}-ACS/I-Band \\
1.6     & 187370  & 0.539 $\pm$ 0.041   & {\it HST}-NICMOS(NIC2)/H-Band \\
3.6     & 83275.7 & 0.585 $\pm$ 0.003\tna   & {\em Spitzer}/IRAC \\
4.5     & 66620.5 & 1.794 $\pm$ 0.003\tna  & {\em Spitzer}/IRAC \\
5.8     & 51688.4 & 3.163 $\pm$ 0.006\tna  & {\it Spitzer}/IRAC \\
8.0     & 37474.1 & 4.589 $\pm$ 0.006\tna  & {\it Spitzer}/IRAC \\
2152.82 & 139.256 & 0.799 $\pm$ 0.082   & PdBI \\
61414   & 4.8815  & 0.866 $\pm$ 0.027   & VLA \\
\cutinhead{Background Galaxy RXJ1131 (deblended bands)}
0.555   & 540167  & 0.009 $\pm$ 0.004\tnb  & {\it HST}-ACS/V-Band \\
0.814   & 368295  & 0.041 $\pm$ 0.005\tnb  & {\it HST}-ACS/I-Band \\
1.6     & 187370  & 0.133 $\pm$ 0.004\tnb   & {\it HST}-NICMOS(NIC2)/H-Band \\
3.6     & 83275.7 & 5.034 $\pm$ 0.002  & {\em Spitzer}/IRAC \\
4.5     & 66620.5 & 6.009 $\pm$ 0.002  & {\em Spitzer}/IRAC \\
5.8     & 51688.4 & 7.557 $\pm$ 0.003   & {\it Spitzer}/IRAC \\
8.0     & 37474.1 & 9.881 $\pm$ 0.004  & {\it Spitzer}/IRAC\\
2152.82 & 139.256 & 0.39 $\pm$ 0.12\tnc   & PdBI \\
61414   & 4.8815  & 1.273 $\pm$ 0.042   & VLA
\enddata
\label{tab:photometry}
\tablecomments{The IRAC photometry for channel 1 (3.6\,$\micron$) is extracted directly from the image and
from the Spitzer Heritage Archive for channels 2$-$4 (4.5, 5.8, and 8.0\,$\micron$). The flux uncertainties quoted for radio and mm observations (PdBI, CARMA, and VLA) do not include those from absolute flux calibration.
All upper limits are 3$\sigma$.}
\tablenotetext{a}{Flux obtained using aperture photometry after subtracting the emission of RXJ1131 from the total emission.}
\tablenotetext{b}{A contribution from the quasar has been removed (see C06), and thus the flux density corresponds to the host galaxy only.}
\tablenotetext{c}{Flux extracted from the residual map after subtracting a point-source model. For SED modeling, we use $S_{\rm 2mm}$\eq0.27\pmm0.08\,mJy to exclude synchrotron emission (see \Sec{deblend}).}
\tablerefs{The {\it HST} photometry is adopted from C06.}
\end{deluxetable}

\section{Analysis} \label{sec:anal}

\subsection{Lens Modeling} \label{sec:lensmodel}
\begin{figure}[!htbp]
\includegraphics[width=0.45\textwidth]{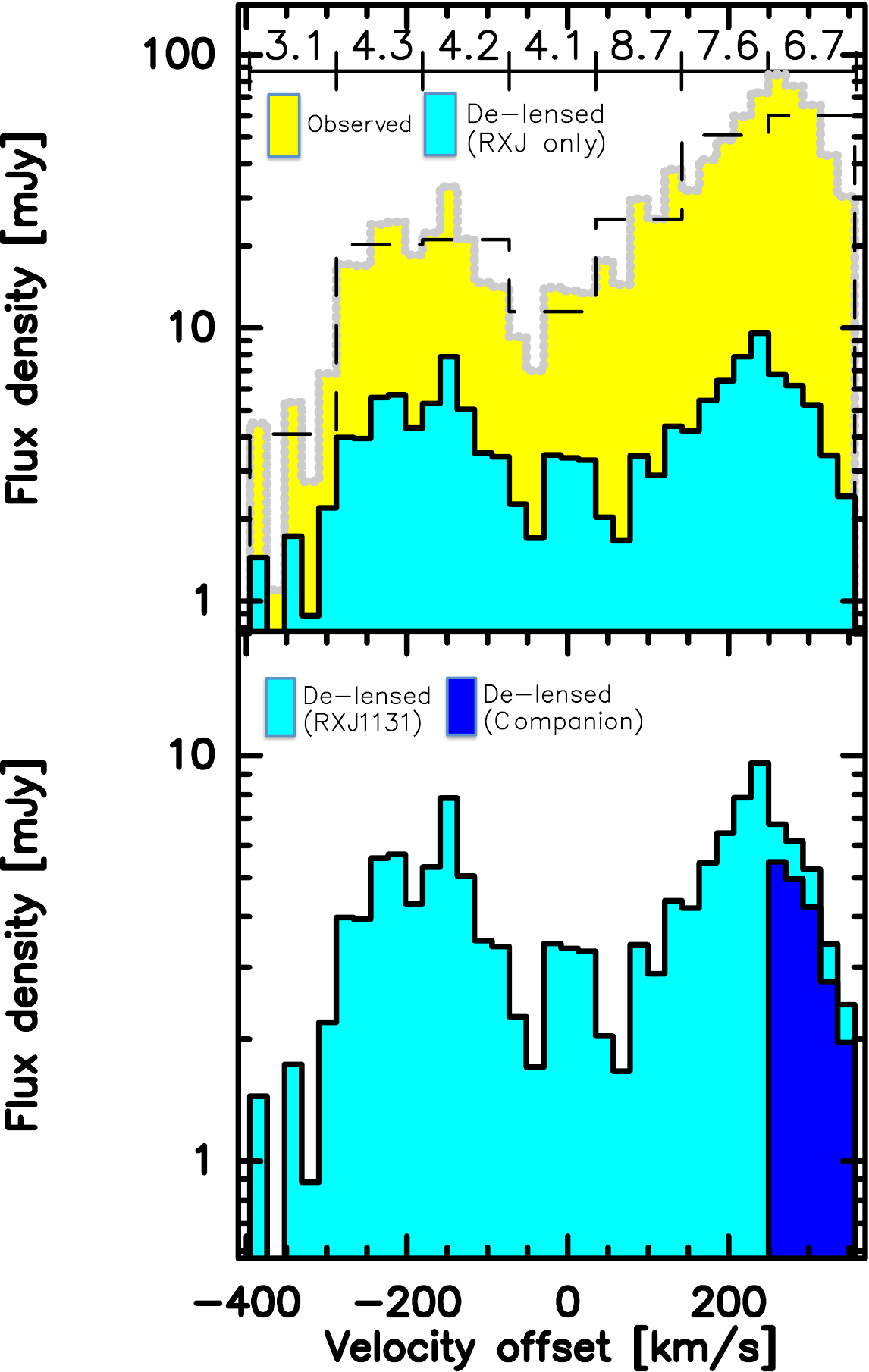}
\caption{Top: the full resolution \bco spectrum (yellow histogram) and
the binned spectrum (dashed line) with the seven $\Delta v$\,$\sim$\,105\,\kms channels used for lens modeling. The
light blue histogram shows the ``intrinsic'' line profile of RXJ1131
after subtracting a contribution from its companion galaxy and
correcting for lensing using the magnification factor $\mu_{\rm L}$ as annotated by the horizontal bar shown
above each respective model channel.
Bottom: the ``intrinsic'' line profile of RXJ1131 (light blue) and of its companion (dark blue).
The y-axes are shown on a log scale.
\label{fig:delensed}}
\end{figure}

At the angular resolution of the \bco data, the source is resolved
$\gtrsim$2 resolution elements.
Given the extent of the lensed emission (see \Fig{CO21mom}),
this implies that we do not resolve
structures (e.g. knots and arcs) of the lensed emission
in our \bco data.
Nevertheless, the high spectral
resolution of these data provides kinematic information on
spatial scales smaller than the beam (see \Fig{CO21mom}).
Hence, we reconstruct the intrinsic line profile and source-plane velocity structure
by carrying out a parametric lens modeling over different
channel slices of the interferometric data using our lensing code
\uvmcmcfit
(\citealt{uvmcmcfit15a}; see \citealt{Bussmann15a} for details of the code).
Our approach follows a similar strategy as \citet{Riechers08a}, who reconstruct a source-plane
velocity gradient and constrain the gas dynamics in the $z$\,$>$\,4 quasar host galaxy of
PSS J2322$+$1944,
which is also lensed into an Einstein ring configuration.
To ensure adequate SNRs for lens modeling, we bin the frequency channels by a factor of five
to produce seven independent $\Delta v$\,$\sim$\,105\,\kms channels (dashed line in \Fig{delensed})
that cover the full linewidth of $\sim$750\,\kms.
\begin{deluxetable}{lcc}[htbp]
\tabletypesize{\scriptsize}
\tablecolumns{3}
\tablecaption{Lens parameters constrained by models of seven velocity channels}
\tablehead{
\multicolumn{2}{c}{Parameters} &
\colhead{Median values} 
}
\startdata
Offset in RA    & (\arcsec)   &  0.004$\pm$0.027\\
Offset in Dec    & (\arcsec)   & 0.003$\pm$0.027\\
Axial Ratio      &             & 0.56$\pm$0.16\\
Position Angle   & (deg)       & 103$\pm$22\\
Einstein Radius\tna  & (\arcsec)   & 1.833$\pm$0.002
\enddata
\label{tab:lens}
\tablenotetext{a}{This corresponds to mass of $M(\theta$\,$<$\,$\theta_{\rm E})$\eq(7.42\pmm0.02)\E{11}\Msun within the Einstein radius.}
\tablecomments{Parameters describing the foreground lens are
obtained based on the median in the preliminary models (see text for details).
All angular offsets are with respect to
$\alpha$\,=\,11$^{\rm h}$31$^{\rm m}$51\fs44,
$\delta$\,=\,$-$12\degr31\arcmin58\farcs3 (J2000).}
\end{deluxetable}

We model the lens mass distribution using a singular isothermal
ellipsoid (SIE) profile, which is described by five free parameters: the
positional offset in R.A. and Dec. relative to an arbitrary chosen
fixed coordinate in the image, the Einstein radius, the axial ratio, and the
position angle.
Positional offset between the foreground galaxy and the pre-defined coordinate is initialized
using the VLA radio continuum map.
We impose a
uniform prior of $\pm$0\farcs05 in both $\Delta$R.A. and $\Delta$Dec.,
motivated by the astrometry uncertainties in the VLA image as well as
the uncertainties provided by previous SIE lens model \citepalias{Claeskens06a}.
We initialize the Einstein radius based on the model parameters reported by \citetalias{Claeskens06a}
and impose a uniform prior using $\pm$3$\sigma$ of their uncertainties.
The sources are modeled using elliptical Gaussian profiles, which are
parameterized by six free parameters: the positional offset in R.A.
and Dec. relative to the lens, the intrinsic flux density, the effective
radius, the axial ratio, and the position angle. The position of each source
is allowed to vary between $\pm$1\farcs5 (i.e., within the Einstein radius)
and the effective radius is allowed to vary from 0\farcs01$-$2$^{\prime\prime}$.

Our code uses an Markov Chain Monte Carlo (MCMC) approach to sample the
posterior probability distribution function (PDF) of the model parameters.
In each model, we require a target acceptance rate of $\sim$0.25$-$0.5
and check for chain convergence by inspecting trace plots
and by requiring that the samples are obtained beyond at least an autocorrelation time.
We thus employ $\sim$50,000 samples as the initial ``burn-in'' phase
to stabilize the Markov chains (which we then discard) and
use the final $\sim$5,000 steps, sampled by 128 walkers, to identify
the posterior. Here, we
identify the best-fit model and the quoted uncertainties using the
median and the 68\% confidence intervals in the marginal PDFs.

We first obtain a preliminary lens model for each channel slice independently,
where their lens parameters are allowed to vary and are initialized according
to the aforementioned way. We obtain the final model
by repeating the modeling over each slice but fixing their lens parameters
to the overall median in the preliminary models,
as listed in \Tab{lens}.
This ensures that all models share the same lens profile.
The magnification factors in \Tab{model} are determined by taking the ratio
between the image plane flux and the source plane flux of each model.

Our model parameters in \Tab{lens}, describing
the mass distribution of the lensing galaxy, are consistent (within the uncertainties)
with that of the SIE model presented by \citetalias{Claeskens06a}. We find a mass of
$M(\theta$\,\,$<$\,\,$\theta_\textrm{E})$\,=\,(7.47\,$\pm$\,0.02)\,$\times$\,10$^{11}$\,\Msun
within the Einstein radius.

\subsubsection{Interpretation of the Source-plane Morphology} \label{sec:caveat}
\begin{figure*}[tbph]
\centering
\begin{tabular}{c}
\includegraphics[trim=0 0 0 0, clip, width=1.0\textwidth]{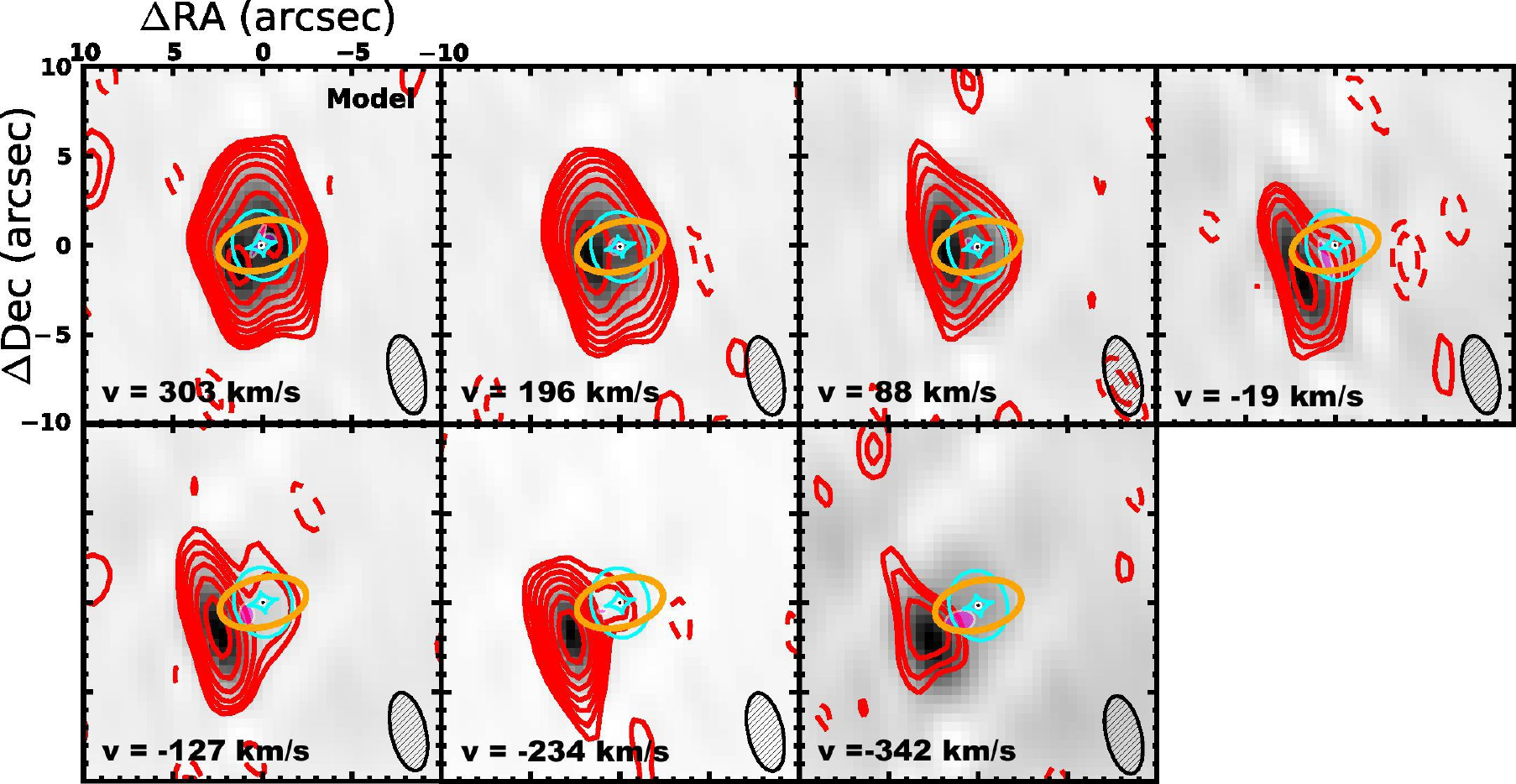} \\
\includegraphics[width=0.98\textwidth]{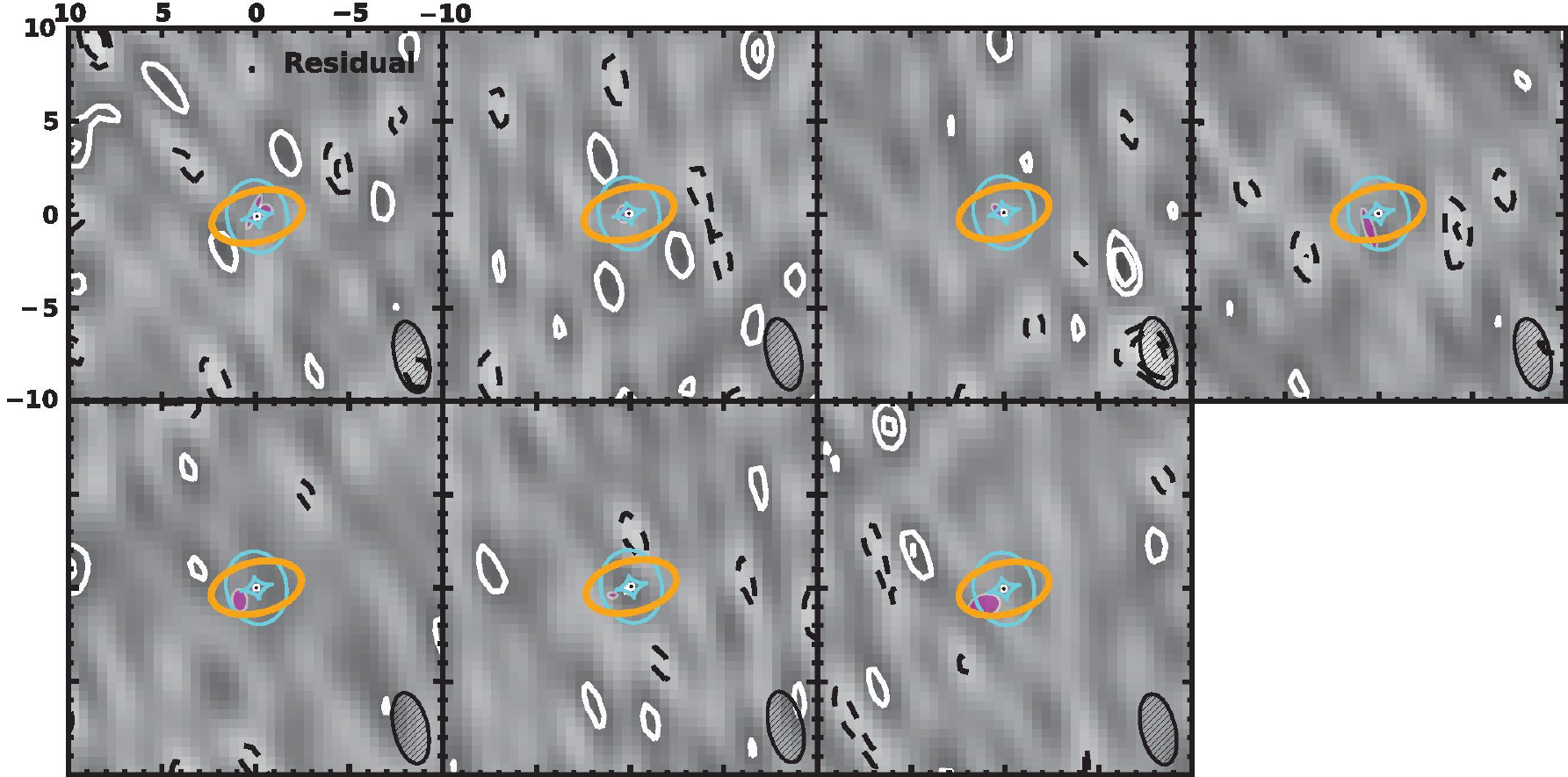}
\end{tabular}
\caption{Best-fit lens models of the PdBI \bco data in different velocity channels, as listed in \Tab{model}.
Top: each panel corresponds to a channel map of width 107.5\,\kms centered on the indicated velocity.
The observed emission (red contours) is over-plotted atop the best-fit model (grayscale).
Bottom: residual images obtained by
taking a Fourier transform after subtracting the best-fit model from the
data in the $uv$-domain.
In all panels, the location of the foreground lensing galaxy is indicated by a black dot and
its critical curve is traced by the orange solid line; locations and
morphologies (half-light radii) of the reconstructed sources are
represented by magenta ellipses; the caustic curves are represented as cyan lines.
Contours start
at $\pm$3$\sigma$ and increment in steps of 3$\times$2$^n\sigma$,
where $n$ is a positive integer.
The beam of the PdBI observations is shown in the bottom right corner.
The reconstructed source-plane positions, as represented by the magenta ellipses, demonstrate an intrinsic velocity gradient of the \bco emission in RXJ1131.
The best-fit model of the red-most channel (top left panel) contains two source components --- RXJ1131 and its companion galaxy.
\label{fig:model}}
\end{figure*}

The reconstructed source locations, as represented by magenta ellipses in \Fig{model}, demonstrate
an intrinsic velocity gradient across the source plane, which is
consistent with a kinematically-ordered disk-like galaxy.
Additional support to the disk conjecture
can be found in the double-horned line profile (\Fig{CO21spec})
and the observed (image plane) velocity field (\Fig{CO21mom}). Furthermore,
\citetalias{Claeskens06a} also find that the reconstructed source plane emission in optical-NIR
is best-reproduced using a $n$\,=\,1 Sersic profile.
We thus interpret RXJ1131 as a disk galaxy.

A better fit is found for the lens model of
the red-most channel if we add a second source component (see
top left panel in \Fig{model}). This is consistent with previous results
reported by \citet[hereafter B08]{Brewer08a}, who find an optically faint companion
(component F in their paper) $\sim$2.4\,kpc in projection from the AGN host galaxy in $V$-band,
and with \citetalias{Claeskens06a}, who find evidence for an interacting galaxy near RXJ1131.
Spatially, the red velocity component of the CO emission
also consistent with this component F. It is therefore likely that we
detect \bco emission in a companion galaxy.
\defcitealias{Brewer08a}{B08}

We decompose the total line flux into two components:
one from RXJ1131 and the other from its companion.
Since the companion is only detected in the red-most channel, we
derive its intrinsic gas mass using the best-fit flux
densities and magnification factors obtained from the models of this channel.
Assuming a brightness temperature ratio
of $r_{\rm 21}$\,=\,1 between \bco and \aco lines and
a CO luminosity-to-H$_2$ mass conversion factor of
\alphaco\,=\,0.8\,\alphaU, we find
a molecular gas mass of $M_{\rm gas}$\,=\,$($1.92\pmm0.09$)$\,\E{9} \Msun.
For the molecular gas mass in RXJ1131, we derive
its intrinsic line flux over the FWZI linewidth
using the respective magnification
factors listed in \Tab{model}, which to
first order takes into account the effect of differential lensing.
This yields $I_{\small \bco}$\,=\,2.93\pmm0.70 Jy\,\kms,
where the uncertainty includes those on
the magnification factors.
Adopting the same brightness temperature ratio and \alphaco\ as
used for the companion, this corresponds to a gas mass of
$M_{\rm gas}$\,=\,$($1.38\pmm0.33$)$\,\E{10} \Msun, which
implies a {\em gas} mass ratio of $\sim$7:1 between RXJ1131 and its companion.

The spatial resolution of the data in hand
is a few arcsec, which implies that despite the high SNR and spectral
resolution, constraints on the intrinsic sizes of the lensed galaxies are modest, and thus the magnification
factors may be under-predicted \citep[see e.g.,][]{Bussmann15a,Dye15a,Rybak15a}.

\subsubsection{Spatial Extent and Differential Lensing} \label{sec:differential}
\begin{figure*}[!htbp]
\centering
\includegraphics[width=1.0\textwidth]{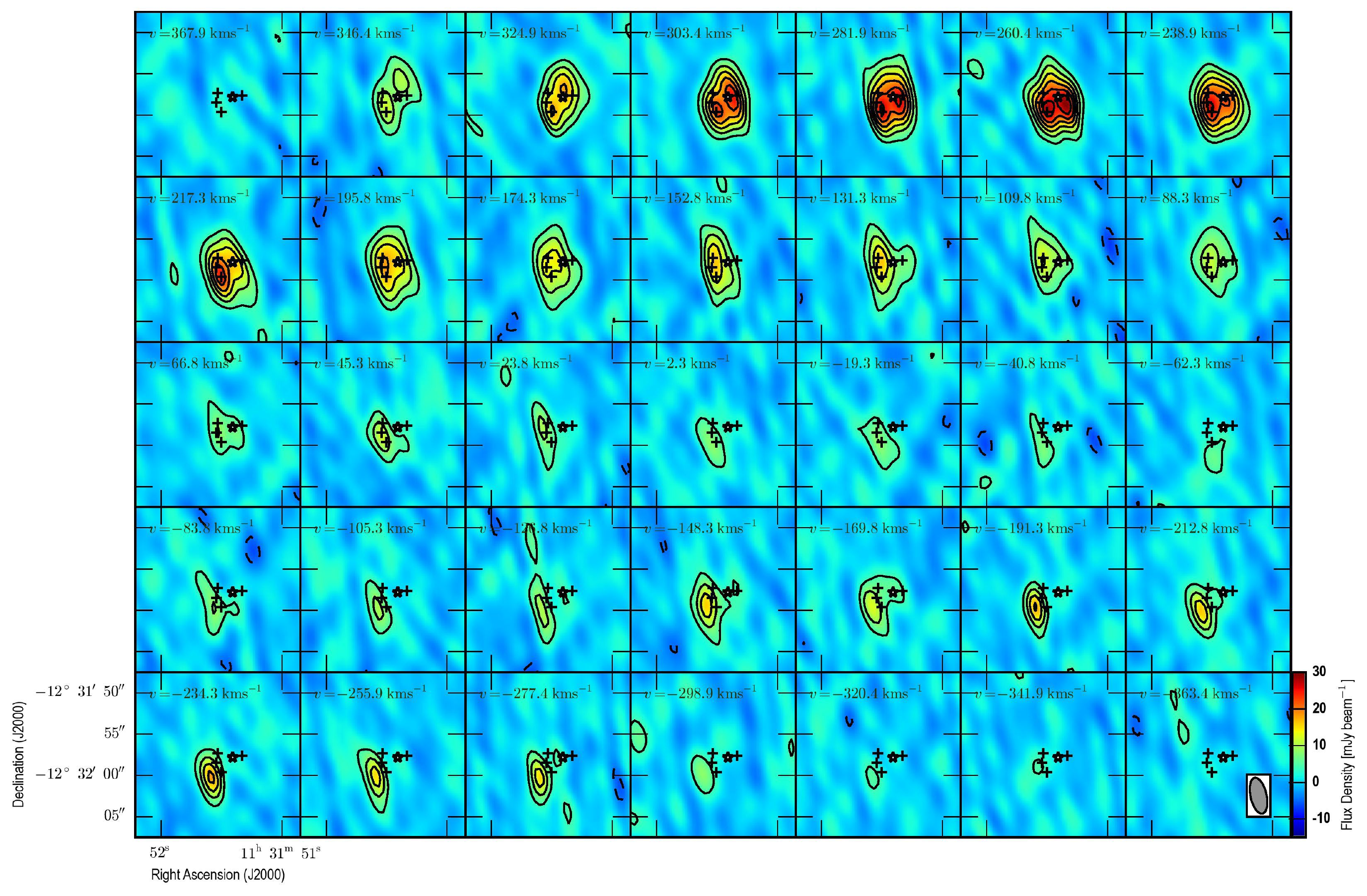}
\caption{
Channel maps of the PdBI \bco data at 22\,\kms resolution.
Black crosses indicate the positions of the lensed knots (AGN emission,
which correspond to components ABCD in \citetalias{Claeskens06a}). The white-filled
star indicates the position of the foreground lensing galaxy (component G
in \citetalias{Claeskens06a}).
Central velocities are shown at the top of each map.
Contours start and increment in steps of
$\pm$3$\sigma$. The beam is denoted in the bottom right panel. \label{fig:chanmap}}
\end{figure*}

\begin{figure*}[!htbp]
\centering
\includegraphics[width=0.9\textwidth]{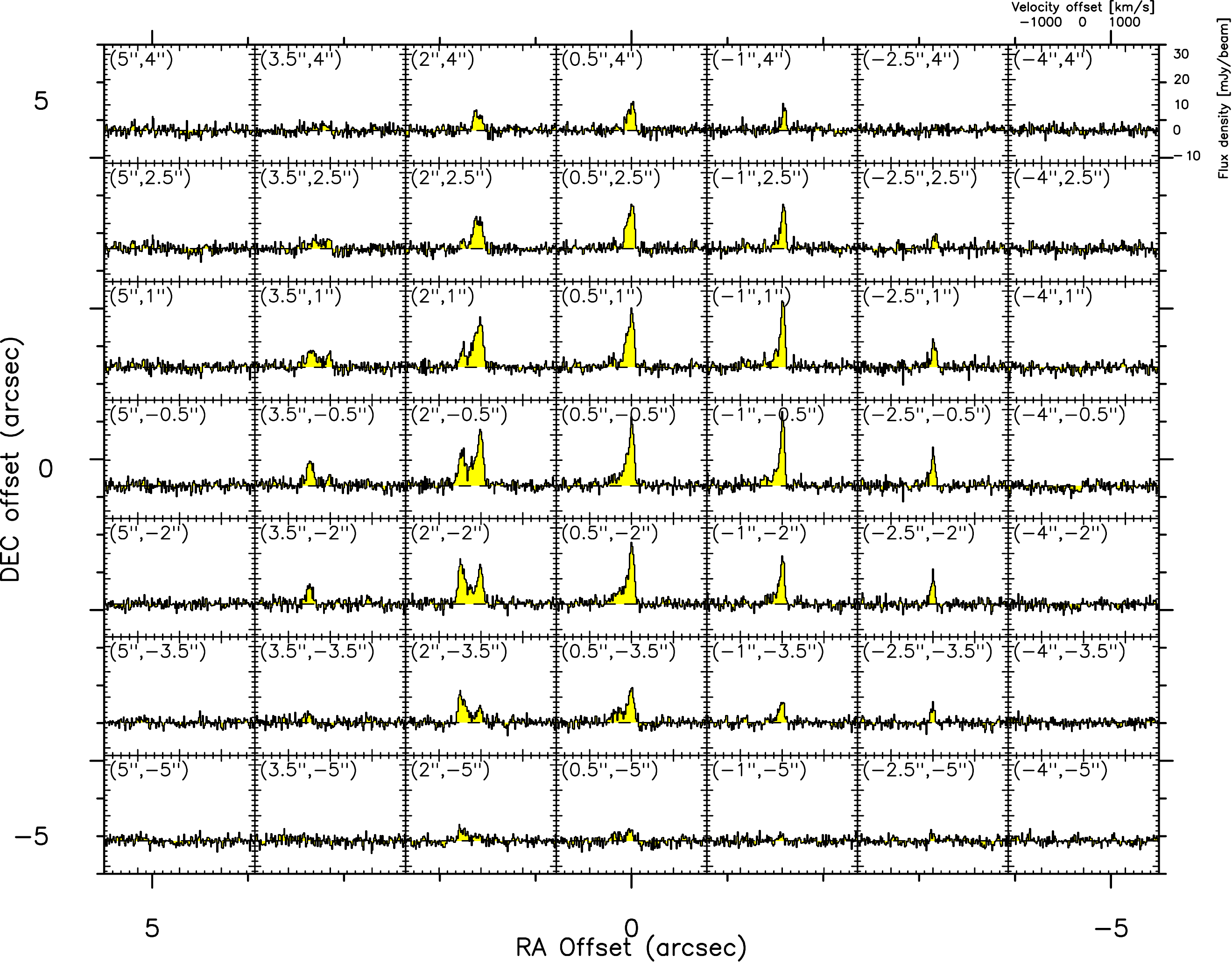}
\caption{
\bco spectrum as a function of position, binned by 3 pixels in each
direction (1\farcs5).
The spectra map covers an extent of $\sim$10''$\times$10''
centered on the pixel that corresponds to coordinates of the lensing galaxy ($\alpha_{\rm J2000}$\,=\,11$^{\rm h}$31$^{\rm m}$51\fs44, $\delta_{\rm J2000}$\,=\,$-$12\degr31\arcmin58\farcs3).
Spatial offset in arcsec is denoted in top left corner of each panel.
The velocity and flux density scales are denoted in the top right panel.
\label{fig:spatialSpec}}
\end{figure*}

In the image-plane integrated line map shown in \Fig{CO21mom}, the redshifted component is
cospatial with the Einstein ring that is seen in the
optical image, with most of its apparent flux originating along the lensing arc,
whereas centroid of the blueshifted emission is offset to the SE of
the lensing arc. This suggests that the CO-emitting region in RXJ1131 is extended.
To further illustrate this, we show the
channel maps of 21.5\,\kms width and a spatial spectra map of 1\farcs5 resolution in
\Fig{chanmap} and \Fig{spatialSpec}, respectively. These figures
show that redshifted emission
is present to the west, peaking toward the lensing arc (black crosses in
\Fig{chanmap}), and shifts to the east with decreasing velocity
(blue wing).
This is consistent with the source plane positions in our models and
is suggestive of an extended CO emitting region.

\begin{deluxetable}{lcc}[!htbp]
\tabletypesize{\scriptsize}
\tablecolumns{3}
\tablecaption{Magnification factors of various kinematic components in \bco}
\tablehead{
\colhead{Velocity Range (\kms)} &
\colhead{Source 1 $\mu_{\rm L}$} &
\colhead{Source 2 $\mu_{\rm L}$}
}
\startdata
346\,$-$\,260       & 6.7 $\pm$ 2.5 &  7.2 $\pm$ 5.6 \\
238\,$-$\,153       & 7.6 $\pm$ 1.6 & \\ [0.5ex]
131\,$-$\,45        & 8.7 $\pm$ 2.0 & \\ [0.5ex]
24\,$-$\,-62    & 4.1 $\pm$ 0.9 & \\ [0.5ex]
-84\,$-$\,-170  & 4.2 $\pm$ 0.6 & \\ [0.5ex]
-191\,$-$\,-277 & 4.3 $\pm$ 2.4 & \\ [0.5ex]
-300\,$-$\,-385 & 3.1 $\pm$ 0.9 & \\ [0.7ex]\hline
weighted average & 4.4 & \rule{0pt}{1\normalbaselineskip}\\ [0.5ex]
median & 5.5 &
\enddata
\label{tab:model}
\tablecomments{First column corresponds to the rest-frame velocity ranges taken from the center of an unbinned channel
(see \Fig{delensed}). Each row corresponds to a (binned) channel slice used for
lens modeling. Source 1 is RXJ1131 and source 2 is its companion. }
\end{deluxetable}

Previous studies of RXJ1131 find evidence for differential lensing across
the {\it HST} $V$-, $I$-, and $H$-bands, where the
magnification factor varies from 10.9 to 7.8 \citepalias{Claeskens06a}.
This indicates that the emission from different stellar populations
within the host galaxy have various spatial extents and positions with respect to the caustic.
The best-fit lens models obtained here for different CO channels show that differential lensing also plays
an important role in the observed \bco emission, with a magnification factor $(\mu_{\rm L})$ that varies
from 8.7 to 3.1 across different kinematic components (\Tab{model}).
The asymmetry in the line profile (\Fig{CO21spec} and \Fig{delensed}) is therefore predominantly a result of
the redshifted CO-emitting gas being more strongly-magnified
than the
blueshifted component.
A secondary reason is likely due to
the inclusion of the emission of the companion in the most redshifted velocity channels.
The variation in $\mu_{\rm L}$ found across channels is consistent with the source plane
positions relative to the caustics in \Fig{model}, where the red wing
emission mainly originates near the cusp
of the caustic and the blue wing emission is located beyond the caustics.
In fact, the intrinsic line flux of the redshifted and
blueshifted emission in RXJ1131 (after subtracting a contribution from the companion)
are $I_{\small \bco}$\,=\,1.26\pmm 0.23 Jy\,\kms and 1.25\pmm0.23 Jy\,\kms, respectively,
implying an intrinsically symmetric line profile (\Fig{delensed}). This is consistent with the symmetric source-plane
velocity gradient in our lens model (\Fig{model} and \Fig{PV}).

\subsection{\bco Kinematics}
Fitting two Gaussians with a common FWHM
to the ``intrinsic'' \bco line profile of RXJ1131 (after correcting for lensing using
the magnification factors for various channels and separating the emission from RXJ1131 and its companion),
we find a roughly symmetric double-horned profile with a flux ratio of 1.2\pmm0.4 between the peaks, which
are separated by
$\Delta v_{\rm sep}$ = 387\pmm45\,\kms, and each with a
FWHM of 220\pmm72\,\kms.
The peak separation obtained from this ``intrinsic'' line profile is
slightly lower than that obtained from the observed spectrum (\ie without lensing corrections).
This discrepancy is likely a result of differential lensing, which causes the line peak of the red wing
to shift towards higher velocity channels, and thereby biasing the centroid of
one Gaussian to higher velocity than otherwise.
To facilitate a comparison (\Sec{sizes}) with previous works, which were observed at lower spectral resolution,
we also fit a single-Gaussian to the intrinsic line profiles.
This yields FWHMs of 600\pmm160\,\kms for RXJ1131
and 73\pmm43\,\kms for the companion galaxy.

A clear velocity gradient and a high
velocity dispersion ($\gtrsim$400\,\kms) near the central region
are seen in \Fig{CO21mom}. While beam smearing is inevitably the
dominant factor in the observed velocity dispersion
at the spatial resolution of these data, the exceedingly
high velocity dispersion may hint
at potential perturbations from the AGN, or internal turbulence due to
interactions with the companion, and/or instability due to the large gas
content.
Therefore, in this scenario, RXJ1131 is
consistent with a disrupted disk galaxy hosting an optically
bright quasar and in the process of merging.

\subsection{\bco Dynamical Modeling} \label{sec:dynamics}
\begin{figure*}[!htbp]
\centering
\includegraphics[trim=0 0 30 0, clip, width=0.4\textwidth]{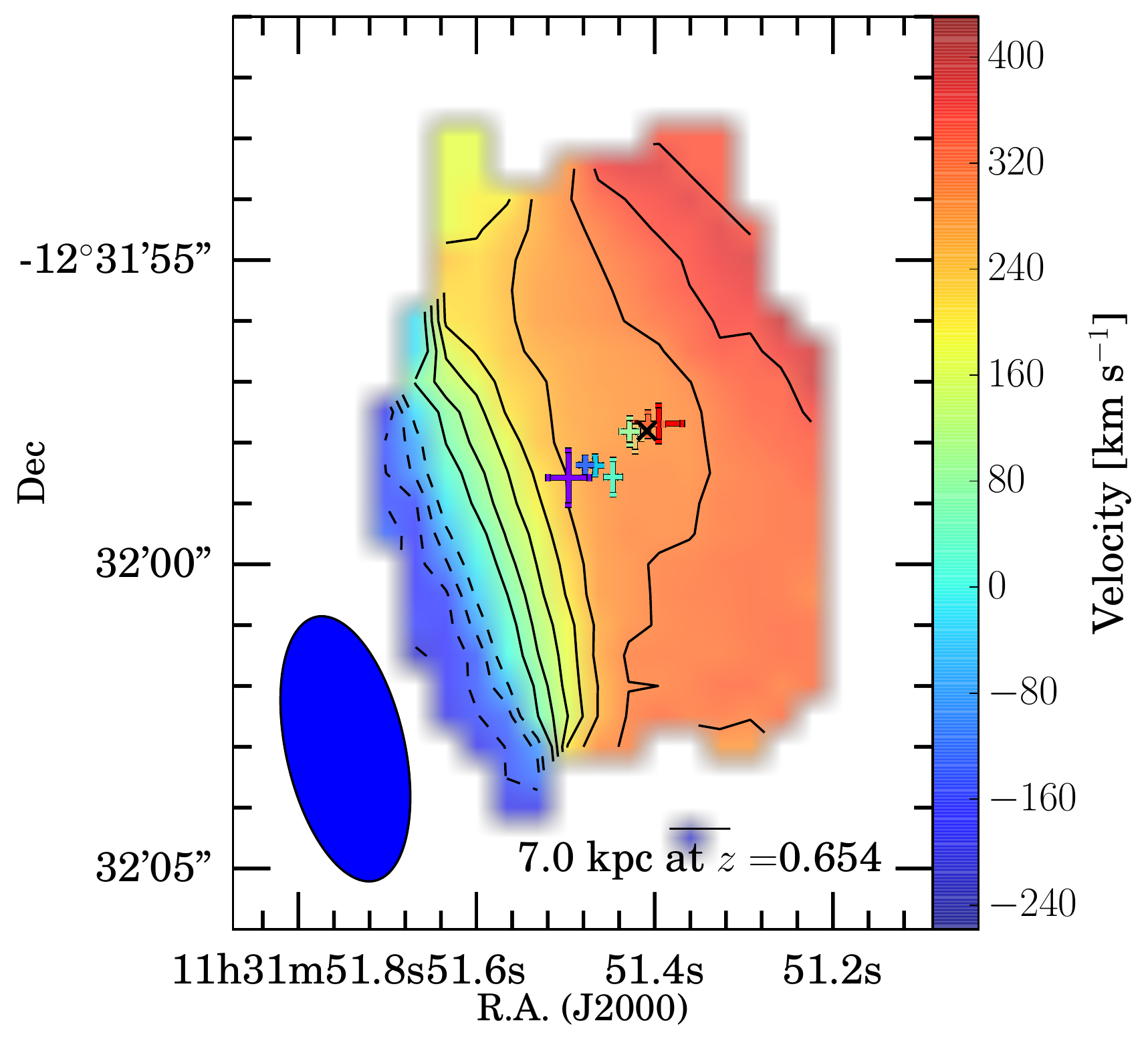}
\includegraphics[width=0.525\textwidth]{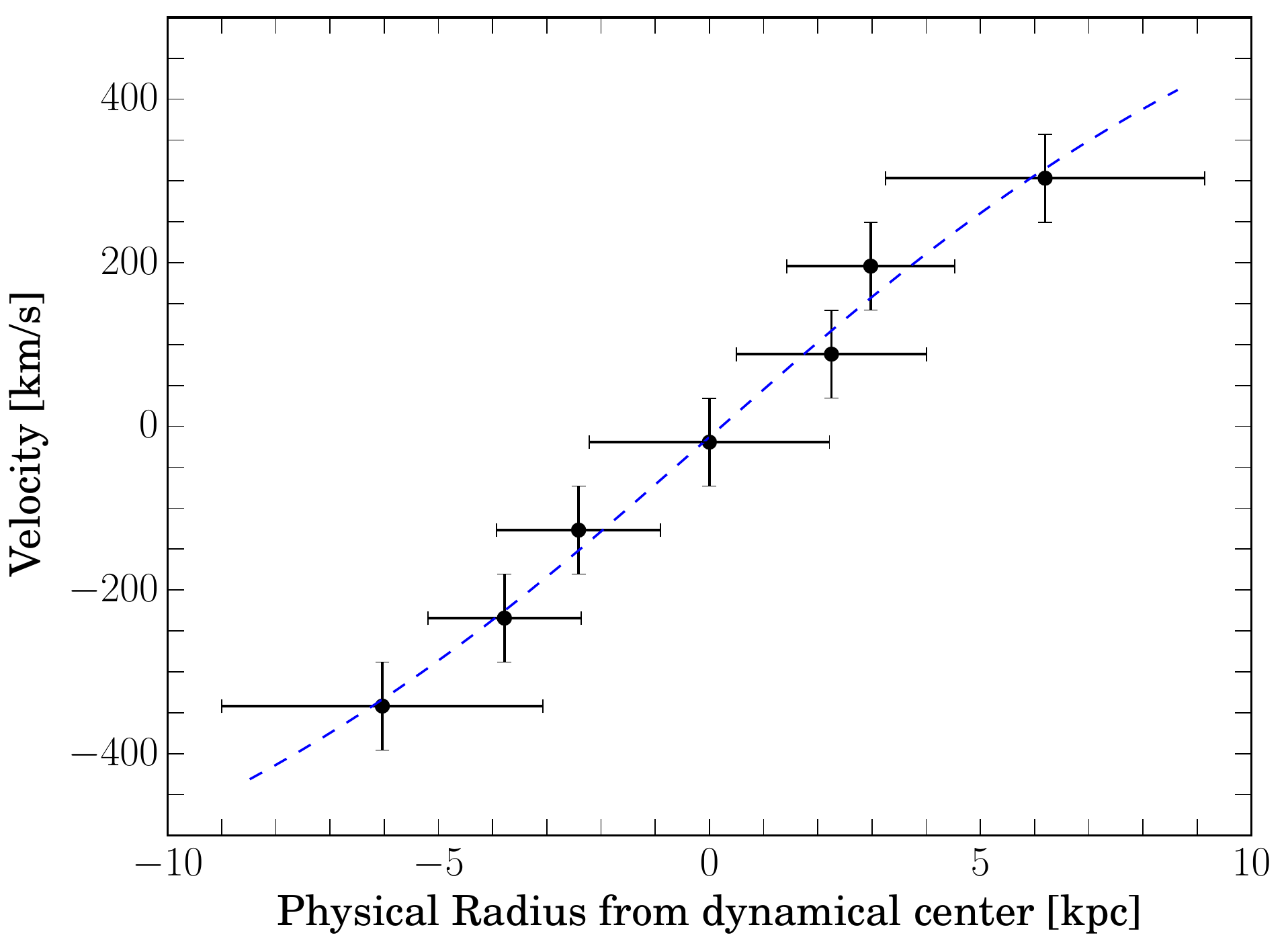}
\caption{Velocity gradient of the \bco emission observed in RXJ1131 and the de-lensed velocity gradient obtained from lens modeling.
Left: source-plane positions obtained from the best-fit lens models (presented in \Sec{lensmodel}) are shown
as markers
atop the observed first moment map (i.e., in the image-plane; see \Fig{CO21mom}).
The markers are color-coded by their centroid velocities.
The contours are in steps of 50\,\kms.
Despite the distorted first moment map in the image-plane due to differential lensing and beam smearing effects, the source-plane velocity gradient reconstructed from dynamical lens modeling suggests that
RXJ1131 is intrinsically a kinematically-ordered disk.
Right: PV slice extracted along the fitted major axis at PA\,=\,121\degr based on
the reconstructed source-plane velocity gradient.
Dashed line shows the best-fit ``rotation curve'' using an arctangent model.
The vertical error bars show the channel width for
each model and the horizontal error bars are the
1$\sigma$ uncertainties on the source-plane positions along the major axis.
\label{fig:PV}}
\end{figure*}

As discussed in \Sec{caveat}, we interpret RXJ1131 as a disk galaxy as it displays
a kinematically-ordered velocity gradient in the source-plane velocity map of the CO emission,
a symmetric double-horned line profile (\Fig{delensed}, \Fig{model} and, left panel of \Fig{PV}),
and a disk-like morphology in the source-plane reconstruction of the optical-NIR emission
\citepalias{Claeskens06a}.
We extract a one dimensional position-velocity (PV) profile
by assuming that the source-plane centroids of different velocity components
obtained from dynamical lens modeling
are dominated by the tangential component of the
true velocity vector of a rotating disk, i.e., each velocity component would
be seen as lying along the major axis of a rotating disk if observed with sufficiently high angular resolution
(see right panel of \Fig{PV}).
In this process, the positions for each velocity component (plotted as data points in the right panel of \Fig{PV})
are extracted along the best-fitted major axis, which is along a PA of 121\degr.

We then attempt to characterize the molecular gas kinematics using an
empirically-motivated disk model \citep[\eg][]{Courteau97a,Puech08a,Miller11a}:
\begin{equation}
V = V_0 + \frac{2}{\pi} V_{a} \arctan(\frac{R}{R_{t}}),
\end{equation}
where $V$ is the observed velocity, $V_0$ is the velocity at dynamical center,
$V_{a}$ is the asymptotic velocity, and $R_{t}$ is the ``turnover''
radius at which the rising part of the curve begins to flatten.
We perform non-linear least squares fitting using an orthogonal distance
regression to find the best-fit parameters,
taking into account the uncertainties in both velocity (channel width) and
distance offset. We also place an upper limit on $R_{t}$\,$<$15 kpc
to keep this parameter physical \citep[\eg][]{Puech08a,Miller11a}.
The parameter uncertainties are inferred based on a Monte Carlo simulation
of 500 iterations, where the input parameters are perturbed
according to random Gaussian distributions with standard deviations
corresponding to their uncertainties.
Using this model, we find $V_{a}$\,=\,988\,$\pm$\,618\,\kms,
$R_{t}$\,=\,10.9\,$\pm$\,7.8\,kpc, and $V_0$\,=\,0\,$\pm$\,9\,\kms.
However, since the emission is not resolved along the flat regime
of the rotation curve, the asymptotic velocity
and the ``turnover'' radius are poorly constrained.
In particular, $V_{a}$ and $R_{t}$ are highly correlated with a
Pearson coefficient $R$\,=\,0.998, and $-$0.400 between $V_{a}$ and $V_0$.

The asymptotic velocity ($V_{a}$) --- an extrapolation of the model
out to radius beyond the disk scale-length and half-light radius ---
is not equivalent to the maximum observed velocity ($V_{\rm max}$),
which is commonly used in literature to parameterize disk rotation.
The arctangent model is most commonly used in studies of the
Tully-Fisher relation, where an extrapolation to V$_{2.2}$ (velocity
at 2.2 disk scale-length or $\sim$1.375 half-light radius,
or $\sim$0.7$R_{\rm opt}$\footnote{Radius enclosing 83\% of the light
distribution.}) is typically adopted
as the rotation velocity ($V_{\rm max}$ in their
terminology), since this corresponds to the radius at which the velocity
of a pure exponential disk peaks \citep{Courteau97b}.
Here, we adopt the maximum {\em observed} velocity
$V_{\rm rot}$\,=\,303\,$\pm$\,55\,\kms at 6\,$\pm$\,3\,kpc
from the
dynamical center as a proxy to the rotation velocity.
This radius corresponds to $\sim$0.6\,$R_e$, where $R_e$ is the half-light
radius $\sim$10.3\,kpc inferred from the {\it HST} $I$-band
lens model (\citetalias{Claeskens06a}; converted to
our cosmology).
We note that the source plane half-light radius varies substantially with
wavelength. In particular, the half-light radius is found to be
$\sim$\,4\,kpc and $\sim$7\,kpc in $V$-band
\citepalias{Brewer08a} and $H$-band \citepalias{Claeskens06a}, respectively.
The CO gas is thus of similar spatial
extent as in the $H$ and $I$-bands.

In the rest-frame,
emission in the observed-frame $H$-band corresponds to NIR emission $(\sim$1\,$\micron)$,
tracing radiation from the accretion disk surrounding
the central AGN and also from old and evolved stellar populations;
$I$-band corresponds to roughly the optical $V$-band, tracing stellar radiation from
existing, less massive (\ie longer-lasting) stars;
$V$-band corresponds to roughly $U$-band,  tracing radiation from massive young stars
in the host galaxy. Hence,
the relative compactness observed in the $V$-band may be explained in part
due to the fact that the emission in this band is
more susceptible to dust extinction than in other bands and/or dominated by
a central starburst caused by higher
concentrations of star-forming gas towards the central regions --- owing to
gravitational perturbations induced
from interactions with the companion
\citep[\eg][]{DiMatteo05a}.
This would be consistent with the picture that old stars form first and constitute the bulge component
of a spiral galaxy, and that nuclear starbursts (in the inner few kpc) can be triggered
at a later time as the progenitor disk galaxy interacts with other galaxies to form a larger bulge.

\subsection{Dynamical Mass}\label{sec:dyn}
Assuming the gas to be virialized,
the dynamical mass can be approximated by
$M_{\rm dyn}$\ssim$\sigma^2 R / G$,
where $\sigma$ is the velocity dispersion, or the rotational velocity in the case of a rotating disk model
$($\ie $\sigma$ = $V_{\rm rot}\, \sin\, i)$.
Using a rotational velocity $V_{\rm rot} \, \sin\, i$\,=\,303\,\kms (see \Sec{dynamics}),
we find a dynamical mass of
$M_{\rm dyn}$\,$\sin^2 i$\,$(<$\,6\,kpc$)$ = 1.3\E{11}\,\Msun enclosed
within the CO-emitting region in RXJ1131.
If we instead consider the
\bco line peak separation $(\Delta v_{\rm sep}/2 \sim$200\,\kms$)$ as the rotation velocity, we find
$M_{\rm dyn}$\,$\sin^2 i$\,$(<$\,6\,kpc$)$ = 5.8\E{10}\,\Msun.
We derive an inclination angle of 56.4$\degr$ from the
morphological axial ratio of $a/b\sim$1\farcs8$/$3\farcs25, which we estimate
from the source-plane image reconstructed by \citetalias{Claeskens06a} (Figure 3 in their paper).
This corresponds to an inclination-corrected dynamical mass of
8.3\E{10}\Msun$<$\,$M_{\rm dyn}$\,$<$\,25\E{10}\Msun.
Our estimate should be considered at best an upper limit since
the gas in RXJ1131 is unlikely to be virialized.
In the following sections, we use the
lower limit (8.3\pmm1.9)\E{10}\,\Msun as the dynamical mass as it is
derived in a manner similar to what is commonly used in literature
(\eg \citealt[hereafter S97]{Solomon97a}; \citealt[hereafter DS98]{Downes98a}; \citealt[hereafter G05]{Greve05a}).
\defcitealias{Downes98a}{DS98}
\defcitealias{Solomon97a}{S97}\defcitealias{Greve05a}{G05}\defcitealias{SV05a}{SV05}

Using the velocity dispersion ($\sigma$ = 30\,\kms) obtained by fitting a single Gaussian to the
de-lensed line profile of the companion and a
half-light radius of $R_{\rm CO}$\,=\,4.2\pmm2.8\,kpc from the best-fit lens model,
we find a dynamical mass of
$M_{\rm dyn}$\,=\,(3.5\pmm2.3)\E{9}\,\Msun for the companion,
assuming an inclination angle of $i$\eq30$\degr$.
The uncertainty here only includes that of the CO source size.
On the other hand,
we find $M_{\rm dyn}$\,$\sin^2 30\degr$\,=\,5.8\E{8}\,\Msun
if we adopt the better-constrained $V$-band source size of $\sim$700\,pc \citepalias{Brewer08a}.
Since the $V$-band based dynamical mass measurement is substantially lower than the gas mass,
the $V$-band emitting region may appear to be much smaller than its true extent due
to dust obscuration.

The CO-based dynamical mass estimates correspond to a mass ratio of $\sim$24:1
between RXJ1131 and the companion, with a gas mass ratio of $\sim$7:1 derived in  \Sec{caveat}.
We thus classify the system as a {\em gas-rich,``wet'' minor merger}.

\subsection{SED Modeling} \label{sec:SED}
We fit dust SED models to the 24\,\micron$-$2.2\,mm photometry
using a modified-blackbody (MBB)
function with a power-law attached to the Wien side to account for the MIR excess due to
emission of warm and small dust grains.
The IRAS 60\,$\micron$ and 100\,$\micron$ upper limits are included to constrain the dust peak.
Here, we use a flux density of $S_{\rm 2mm}$\eq0.27\pmm0.08 mJy derived in \Sec{deblend}
instead of the deblended flux listed in \Tab{photometry}
to exclude a potential contribution due to synchrotron emission (see \Sec{deblend}) in the dust SED modeling.
An uncertainty from absolute flux calibration of $\sim$15\% is added in quadrature
to the PdBI 2\,mm continuum photometry in our fitting procedure.

\begin{figure}[!htbp]
\centering
\includegraphics[trim=5 5 5 5, clip, width=0.445\textwidth]{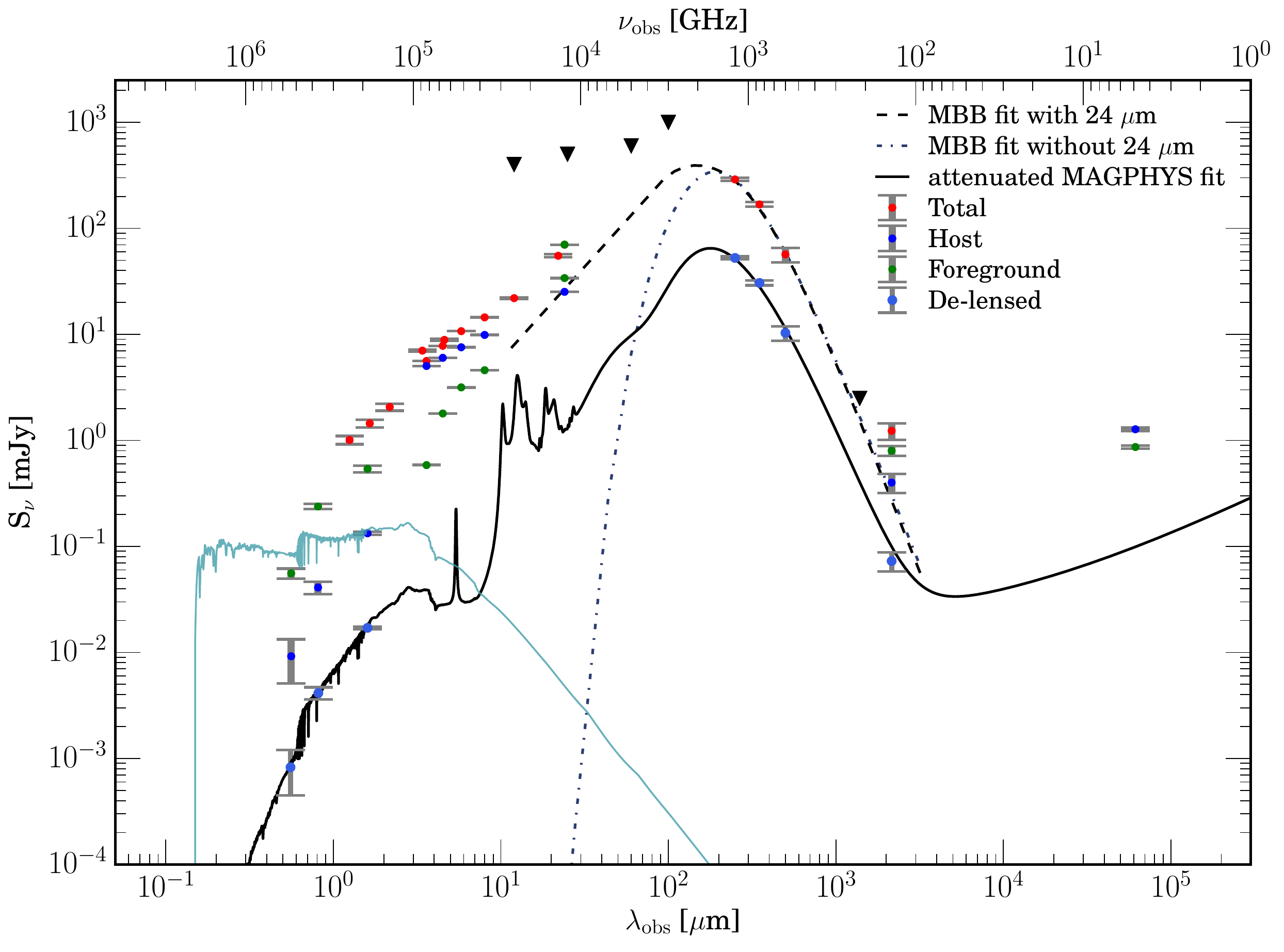}
\caption{SEDs of RXJ1131 at
$z_{\rm CO}$\eq0.654 and the lensing galaxy at $z$\eq0.295.
The photometry data (colored markers) are listed in \Tab{photometry}, except for those
shown in light blue circles, which are corrected for lensing magnification (see \Sec{SED}).
Assuming an MBB$+$power-law model for the thermal dust emission towards RXJ1131,
the dashed (dashed-dotted) line corresponds to the best-fit model
with (without) MIR constraint at 24\,$\micron$.
The solid black line shows the best-fit SED model obtained using the \ncode{magphys} code.
The light blue solid line shows the unattenuated stellar emission.
\label{fig:SED}}
\end{figure}

The fit is performed using the code
\ncode{mbb\_emcee} \citep[e.g.,][]{Riechers13a,Dowell14a}, which samples the posterior
distributions using an MCMC approach and uses instrumental
response curves to perform color correction.
The model is described by five free parameters: the rest-frame characteristic dust
temperature ($T_{d}$), the emissivity index ($\beta$), the power-law index
($\alpha$), the flux normalization at 500\,$\micron$ ($f_{\rm norm}$), and
the observed-frame wavelength at which the emission
becomes optically thick ($\lambda_{0}$). We impose
a uniform prior with an upper limit of 100\,K on $T_d$ \citep[see e.g.,][]{Sajina12a},
a Gaussian prior centered around
1.9 with a standard deviation of 0.3 on $\beta$, and a uniform prior with an upper limit of
1000\,$\micron$ on $\lambda_0$.
We check for chain convergence by requiring that the autocorrelation
length of each parameter is less than the number of steps
taken for the burn-in phase (which are then discarded).
Here we report the statistical means
and the 68\% confidence intervals in the marginal PDFs
as the best-fit parameters, as listed in \Tab{SED}.
The best-fit models are shown in \Fig{SED} along with the broadband photometry that is listed in \Tab{photometry}.

\begin{deluxetable}{lccc}[!htbp]
\tabletypesize{\scriptsize}
\tablecolumns{4}
\tablecaption{SED fitting results}
\tablehead{
\multicolumn{2}{c}{Parameters}      &
\colhead{With 24\micron} &
\colhead{Without 24\micron}
}
\startdata
$T_d$                           & (K)                & 54\petm{8}{10}   &
55\petm{21}{20}  \\ [1.05ex]
$\beta$                         &                    & 1.6\petm{0.5}{0.4}    & 2.2\petm{0.3}{0.3}  \\ [1.05ex]
$\alpha$                        &                    & 1.6\petm{0.5}{0.6}    & 8.5\petm{7.0}{6.2}   \\ [1.05ex]
$\lambda_0$\tna                 & ($\micron$)        & 559\petm{278}{324}    & 365\petm{111}{120}  \\ [1.05ex]
$\lambda_{\rm peak}$\tnb        & ($\micron$)        & 159\petm{19}{40}      & 155\petm{38}{43}  \\ [1.05ex]
$f_{\rm norm,\ 500\micron}$\tnc & (mJy)              & 55\petm{13}{13} & 59\petm{6}{6} \\ [1.05ex]
\LFIR\tnd                       & (10$^{12}$\,\Lsun) & 3.81\petm{1.92}{1.97} & 4.24\petm{2.17}{2.00}      \\ [1.05ex]
$M_{\rm d}$\tne                 & (10$^8$\,\Msun)    & 16\petm{5}{12}        & 14\petm{5}{7}
\enddata
\label{tab:SED}
\tablenotetext{a}{Observed-frame wavelength where $\tau_\nu$\,=\,1}
\tablenotetext{b}{Observed-frame wavelength of the SED peak}
\tablenotetext{c}{Observed-frame flux density at 500 $\micron$}
\tablenotetext{d}{Rest-frame 42.5$-$122.5\,$\micron$ luminosity}
\tablenotetext{e}{Derived assuming an absorption mass coefficient of $\kappa$\eq2.64\,m$^2$\,kg$^{-1}$ at $\lambda$\,=\,125.0\,$\micron$ \citep{Dunne03a}}
\tablecomments{Errors reported here are $\pm$1$\sigma$.
\LFIR and $M_{\rm d}$ are not corrected for lensing.}
\end{deluxetable}

In the first model, we attempt to constrain the power-law index by including
the 24\,$\micron$ data. Based on the resulting posterior PDFs, we find an apparent
IR luminosity (rest-frame 8\,$-$\,1000\,$\micron$) of 8.22\petm{2.75}{2.98}\E{12} \Lsun, a
\fir luminosity (rest-frame 42.5\,$-$\,122.5\,$\micron$) of
3.81\petm{1.92}{1.97}\E{12}\,\Lsun, and a
dust mass of 16\petm{5}{12}\E{8}\,\Msun, none of which are corrected for lensing magnification.
For the mass absorption coefficient, we adopt
$\kappa$\,=\,2.64\,m$^2$kg\pmOne at rest-frame 125.0\,$\micron$
\citep{Dunne03a}.
The dust mass uncertainty does not
include that of the absorption coefficient.

A fit including the MIR 24\,$\micron$ photometry
is likely an upper limit on the \fir luminosity due solely to \SF
in the AGN host galaxy.
If we instead fit for a model excluding this constraint,
two major consequences are immediately apparent.
First, the power-law index is poorly-constrained (see \Tab{SED}).
Second, the steep power-law implies only a small contribution
from the power-law regime
to the total IR luminosity as compared to the graybody component.
Thus, the \fir luminosity in
this model should, in principle, correspond to a
lower limit on the cold dust emission.
Using the best-fit parameters
for this model, we find a total IR luminosity
\LIR (rest-frame 8\,$-$\,1000\,\micron) of 8.67\petm{5.27}{5.27}\E{12}\,\Lsun,
a \fir luminosity \LFIR of 4.24\petm{2.17}{2.00}\E{12}\,\Lsun and a
dust mass $M_{\rm dust}$ of 14\petm{5}{7}\E{8}\,\Msun, all of which are not lensing-corrected.
Taken at face value, this implies an FIR-to-IR luminosity ratio
of $\sim$49\pmm38\%.

The dust temperature from both models is similar to that of
ULIRGs at 0.6\,$<$\,$z$\,$<$\,1.0 (54\,$\pm$\,5\,K;
\citealt[hereafter C13]{Combes13a}).
The \fir luminosity is comparable in both models, which is
not surprising given the lack of constraints in the MIR.
For the subsequent analysis, we adopt the physical quantities
from the first model (\ie with constraints at 24\,\micron).
The choice of SED model does not affect
the derived star formation rate (SFR) given the similar \fir luminosity, and
their dust masses are consistent within the uncertainties.
We correct for lensing using the median magnification
factor $(\mu_{\rm L}\eq5.5)$
from the CO lens models. This yields a
\LFIR of $($6.9\pmm3.6$)$\E{11}\,\Lsun
and
an intrinsic total IR luminosity of $\sim$1.5\E{12}\,$(5.5/\mu_{\rm L})$\,\Lsun, implying that RXJ1131 can be classified as an ULIRG.
Assuming a \citet{Salpeter55a} initial
mass function (IMF), we find a
SFR$_{\rm FIR}$ of 120\pmm63\,\sfrU using a
standard conversion \citep{Kennicutt98a}.
\defcitealias{Combes13a}{C13}

We derive the stellar mass of RXJ1131 by
fitting SED models to the rest-frame UV-to-mm photometry
using the high-$z$ version of the {\sc magphys} code \citep{Magphys08a,Magphys15a}.
Two sets of stellar templates  modeled using either the \citet{BC03a} or the unpublished
Charlot \& Bruzual 2007 stellar population synthesis code
are provided in the {\sc magphys} package.
We adopt the former set.
To minimize contaminations from the quasar, we only fit to the {\it HST}, {\it Herschel}, and PdBI data,
where both the {\it HST} and the PdBI 2\,mm photometry are de-blended from the AGN
(see bottom section of \Tab{photometry}).
The input photometry are corrected for lensing using their respective magnification factors
to account for differential lensing (light blue circles in \Fig{SED}).
We thus find a stellar mass of $M_*$\eq2.95\petm{1.32}{0.86}\E{10}\,\Msun, which
is the median value of the posterior probability distribution and
the uncertainties are derived from the 16$^{\rm th}$ and 84$^{\rm th}$ percentiles.
We note that the models are over-fitted with a best-fit $\chi^2$\eq0.41
which is unsurprising due to sparse sampling of the SED
compare to the number of free parameters.
The resulting dust mass and IR luminosity are consistent with those obtained
from the MBB$+$power-law models within the uncertainties, albeit some differences
in the assumptions behind the two methods.
The consistency may be attributed to the large uncertainties arising from the lack of photometric constraints on the models and the fact that the best-fit parameters from the MBB method are similar to those of the \ncode{magphys} method.

\section{Discussion} \label{sec:diss}
\begin{deluxetable}{lcc}[!htbp]
\tabletypesize{\scriptsize}
\tablecolumns{3}
\tablecaption{Physical properties of RXJ1131 and its companion}
\tablehead{
\colhead{Parameter} &
\colhead{Unit} &
\colhead{Value}
}
\startdata
$r_{\rm 32}$                        &                & 0.78\pmm0.37\\ [0.5ex]
FWHM$_{\rm CO(2-1), RXJ1131}$\tna   & \kms           & 220\pmm72 \\ [0.5ex]
FWHM$_{\rm CO(2-1), RXJ1131}$\tnb   & \kms           & 600\pmm160 \\ [0.5ex]
FWHM$_{\rm CO(2-1), companion}$\tnb & \kms           & 73\pmm43 \\ [0.5ex]
$M_{\rm gas,\ RXJ1131}$             & $10^{10}$\Msun & 1.38\pmm0.33\\[0.5ex]
$M_{\rm gas,\ companion}$           & $10^{9}$\Msun  & 1.92\pmm0.09 \\ [0.5ex]
$R_{\rm CO,\ RXJ1131}$              & kpc            & 6.2\pmm3.0\\ [0.5ex]
$R_{\rm CO,\ companion}$            & kpc            & 4.2\pmm2.8 \\ [0.5ex]
$M_{\rm dyn,\ RXJ1131}$             & $10^{10}$\Msun & 8.3\pmm1.9\tnc \\ [0.5ex]
$M_{\rm dyn,\ companion}$           & $10^{9}$\Msun  & 3.5\pmm2.3\tnc \\[0.5ex]
$f_{\rm gas}$                       & \%             & 18\pmm4\tnd    \\ [0.5ex]
$f_{\rm mol}$                       & \%             & 34\pmm16 \\ [0.5ex]
$L_{\rm IR}$                        & $10^{12}$\Lsun & $\sim$1.5\\ [0.5ex]
$L_{\rm FIR}$                       & $10^{11}$\Lsun & 6.9\pmm3.6\\ [0.5ex]
SFR$_{\rm FIR}$                     & \Msun yr\pmOne & 120\pmm63\\ [0.5ex]
$M_{\rm dust}$                      & $10^{8}$\Msun  & $\sim$3\\ [0.5ex]
GDR                                 &                & 54\pmm13 \\ [0.5ex]
$\tau_{\rm depl}$                   & Myr            & 102\pmm25  \\ [0.5ex]
$M_*$                               & $10^{10}$\Msun & 3.0\pmm1.0  \\ [0.5ex]
$M_{\rm BH}$\tne                    & $10^7$\Msun    & $\sim$\,8 \\ [0.5ex]
$M_{\rm BH}/M_{\rm bulge}$          & \%             & $>$0.27\petm{0.11}{0.08}
\enddata
\label{tab:prop}
\tablecomments{All the parameters have been corrected for lensing magnification. The physical parameters are derived for RXJ1131 and the companion as a single system unless otherwise stated.}
\tablenotetext{a}{From fitting a double Gaussian with a common FWHM to the de-lensed spectrum.}
\tablenotetext{b}{From fitting a single Gaussian to the de-lensed spectrum.}
\tablenotetext{c}{Excluding systematic uncertainties.}
\tablenotetext{d}{Excluding uncertainties in the dynamical masses.}
\tablenotetext{e}{\citet{Sluse12a}.}
\end{deluxetable}

\subsection{ISM Properties} \label{sec:properties}
In this section, we derive the gas properties of the merging system RXJ1131
based on \bco and compare them with those reported by
\citetalias{Combes13a}\footnote{The
\fir luminosity in \citetalias{Combes13a} is derived based on 60\,$\micron$ and 100\,$\micron$ IRAS fluxes,
and using a different definition of
\LFIR: rest-frame 40\,$-$\,500\,\micron. Following this convention,
we find a \fir luminosity of
\LFIR = $($8.8\pmm0.4$)$\E{11}$(\mu_{\rm L}$/5.5$)\pmOne$\,\Lsun and
a SFR of $($150\pmm70$)$\,\sfrU for RXJ1131.} --- the largest sample of CO-detected ULIRGs at similar redshift
(0.6\,$<$\,$z$\,$<$\,1.0).
Their results are based on spatially unresolved \bco and \rot{4}{3} line observations with the
IRAM 30-m single-dish telescope.

\subsubsection{Linewidths and Sizes} \label{sec:sizes}
The FWHM linewidth of $\Delta v_{\rm}$\,$\sim$\,600\pmm160\,\kms found
for RXJ1131 by fitting a single Gaussian
is considerably larger than the statistical average in the \citetalias{Combes13a} sample
(370\,\kms) and
local ULIRGs \citepalias[median: 300\pmm85\,\kms, with the largest being 480\,\kms;][]{Solomon97a}.
Linewidths exceeding 500\,\kms are also commonly observed in
high-$z$ starburst galaxies  \citepalias[e.g.,][]{Greve05a}
and high-$z$ quasar host galaxies \citep[e.g.,][]{Coppin08a},
which are believed to originate from mergers.
The wider CO linewidth observed in RXJ1131 also supports a merger picture.

The CO gas in RXJ1131 is $\sim$6\pmm3\,kpc in radius (in the source plane),
which is more
extended than the average of 3.5\pmm2.3\,kpc in a sample of disk-like U/LIRGs studied by
\citet[]{Ueda14a},
but consistent with their range of 1.1\,$-$\,9.3\,kpc.
Our CO size is also consistent with that of high-$z$
($z$\,$>$\,1) galaxies \citep[$R\sim$\,4$-$20\,kpc;
\citetalias{Greve05a};][]{Daddi10a, Riechers11a, Ivison11a} and
local U/LIRGs in the \citet{Gao99a} sample ($R\lesssim$10\,kpc).

\subsubsection{Gas Mass Fractions and Gas-to-dust Ratio} \label{sec:frac}

We find a dynamical gas mass fraction of $f_{\rm gas}$\eq$M_{\rm gas}$/$M_{\rm dyn}$\eq18\pmm4\%
and a baryonic gas mass fraction of $f_\textrm{mol}$\eq$M_{\rm gas}$/($M_{\rm gas} + M_*$)\eq34\pmm16\% for the merger system (i.e., RXJ1131 and companion).
Recent studies find that the baryonic gas fraction of starburst galaxies has decreased from $f_{\rm mol}$\ssim40\% to $\lesssim$10\% between $z$\ssim2 and $z$\ssim0 \citep[\citetalias{Solomon97a};][]{GS04a, Tacconi06a},
and from $f_\textrm{mol}$\ssim50\% to $\sim$5\% between the same redshift range for
``normal star-forming'' galaxies \citep{Geach11a, Saintonge11a, Tacconi13a}\footnote{These authors use the ``Galactic'' value of
\alphaco\eq4.6\,\alphaU to compute the molecular gas mass.}.
Both the dynamical and baryonic gas mass fractions of RXJ1131+companion are thus consistent with the trend of decreasing molecular gas content since $z$\ssim2
which has been suggested as the cause for the decline in sSFR and cosmic \SF history towards $z$\ssim0 \citep[e.g.,][]{Tacconi13a,CW13,Genzel15a}.

Using the lensing-corrected dust mass, we find a galactic-scale
gas-to-dust ratio (GDR) of
54\pmm13.
This would be higher by a factor of two if we were to adopt a dust mass from the other SED fit that is unconstrained at 24\,$\micron$.
This GDR is lower than the statistical average of 206
in the \citetalias{Combes13a} sample but is well within the broad
range of values measured over their entire sample ($\sim$1$-$770).
Our ratio is also consistent with high-$z$ SMGs
\citep[]{Bothwell13a} and
local ULIRGs \citep{Wilson08a}, but lower than that of the Milky Way by
$\sim 7\sigma$ \citep[ignoring systematic uncertainties;][]{Li01a,Zubko04a,Draine07a}.

There are a number of systematic uncertainties associated with the derived gas-to-dust ratio, in particular
the mass opacity coefficient $\kappa$,
the \alphaco conversion factor, and the brightness temperature ratio $r_{\rm 21}$.
If we instead use the ``Galactic'' \alphaco value, which may be more appropriate for some ULIRGs \citep[\eg][]{Papadopoulos12a} and minor mergers \citep{Narayanan12a},
the gas mass (and thus gas-to-dust ratio) would be $\sim$6 times higher.
We note that this gas mass is physically possible based on the dynamical mass constraints derived in \Sec{dyn}.
On the other hand, we would also obtain a higher gas mass if
we were to assume sub-thermal excitation between \bco and \aco emission.
We also note that the gas-to-dust ratio derived for RXJ1131 may be biased low as the gas is likely to
be more extended than the optically thick dust. Consequently, the overall magnification factor
for the CO gas may be lower than the optically thick dust, which dominates the \fir luminosity.
This would lead to an overestimation of the dust mass
by adopting the CO magnification factor for the dust.

\subsubsection{Star Formation Efficiency and specific SFR}

To first order, the star formation efficiency
$($SFE = \LFIR$/$$M_{\rm gas})$ indicates the \SF rate per unit solar mass of molecular gas available in a galaxy.
Using a wavelength range of 40\,$-$\,500\,$\micron$ defined
in \citetalias{Combes13a} for the far-IR luminosity,
we find an SFE of 58\pmm10 \Lsun $M_{\odot}^{-1}$,
which is on the low end among other U/LIRGs at $z$\,$<$\,0.6
\citep[\citetalias{Solomon97a};][]{Combes11a} but consistent with those of
low-$z$ spiral galaxies \citep[$z$\,$<$\,0.1;][]{SV05a} and high-$z$ disk-like
galaxies, which are also IR luminous galaxies with \LIR$\sim$10$^{12}$\,\Msun \citep{Daddi08a, Daddi10a}.
This suggests that the merger system is converting gas into stars at an efficiency
similar to those of ``normal'' star-forming
disk-like galaxies rather than starburst galaxies
\citep[][\citetalias{Combes13a}]{Tacconi08a, Riechers11a}.
This is in agreement with its disk-like kinematic signatures and its extended molecular gas distribution.
Assuming the \SF continues at the current rate without gas replenishment,
the SFE corresponds to a
gas depletion time of $\tau$\,=\,102\pmm25\,Myr.

The specific star formation rate (sSFR\eq SFR/$M_*$) of 4\petm{2.6}{2.4}\,Gyr\pmOne derived for RXJ1131
is $\lesssim$1.5$\sigma$ above the main sequence according to
the redshift-dependent ``main sequence'' relation in \citet[and references therein]{Tacconi13a}.
Given that RXJ1131 shares similar \SF rate, \SF efficiency, and CO disk size as other ``main sequence'' disk galaxies,
the small elevation in sSFR over the main sequence at $z$\ssim0.7 suggests
that the \SF activity in RXJ1131 may be enhanced by interactions with the companion.

The host galaxy of RXJ1131 is an extended disk with low star formation efficiency in a minor merger
system. Therefore, removal of angular momentum of the gas via gravitational torque is likely inefficient to convert the
entire gas disk into a massive stellar bulge.
In this case, the disk component may be retained upon merging with the companion.
This scenario is consistent with the results from recent simulations, which suggest that bulge formation maybe
suppressed in gas-rich mergers, thereby allowing the formation of large disk galaxies with low bulge-to-disk ratios
\citep{Springel05a, Robertson06a, Hopkins09a}. This also supports the idea that not all mergers will transform into
elliptical galaxies, as in the classical picture \citep{Toomre72a}.

\subsection{Systemic Redshift and Velocity Offset}

\citet[]{Sluse07a} report two sets of AGN lines observed in RXJ1131.
The first set of lines is at $z$\ssim0.654, including the narrow component of the Balmer lines, the \oiii lines, and the \mgii absorption line; the second set is at
$z_{\rm s, QSO}$\ssim0.658, including the broad component of the Balmer lines and the \mgii emission line.
Using the CO line center redshift as the systemic redshift,
we find that the redshift of the first set is fully consistent with the systemic redshift. This
supports previous claims that [O{\scriptsize III}] lines, tracing the narrow line region (NLR),
are good proxies
to the true systemic redshift \citep[e.g.,][]{Vrtilek85a, Nelson00a}.
On the other hand, the second set of lines is redshifted by $\sim$715\,\kms.

Velocity offsets between broad line region (BLR) and NLR lines have been reported in literature.
\citet{Richards02a} find a median offset of
$\sim$100\pmm270\,\kms
between [Mg{\scriptsize II}] and [O{\scriptsize III}] lines in a
sample of $>$3800 quasars,
and \citet{Bonning07a} report
a mean offset of
$\sim$100\pmm210\,\kms
between the broad component of
H$\beta$ and [O{\scriptsize III}] lines in a sample of $\sim$2600 quasars at 0.1$<$$z$$<$0.8,
where only $\lesssim$20 of them (i.e., $<$1\%) are found to have offsets $>$800\,\kms
and $\sim$1\%
are found to have offsets $>$500\,\kms \footnote{\citet{Bonning07a} report the fraction of objects with offset velocities greater
than 500, 800, 1000, 1500, 2000, and 2500\,\kms. We therefore quote the two fractions corresponding to
offset velocities closest
to that of RXJ1131 ($\sim$715\,\kms) in this discussion.}
Thus, large velocity offsets between BLR and NLR lines
comparable to that of RXJ1131 are uncommon but have been observed in some cases.

The observed velocity offset between the BLR and NLR lines may be explained by
a recoiling black hole (BH),
where the BLR is moving at high velocity relative to the bulk of its host galaxy
\citep{Madau04a, Bonning07a, Loeb07a}.
Depending on the initial conditions of the black hole pair (e.g., black hole mass ratio, spin-orbit orientation, spin magnitude),
numerical relativity simulations have shown that recoil velocities can reach up to
$v_{\rm kick}$\ssim4000\,\kms for spinning BHs,
with typical recoil velocities of $v_{\rm kick}$\ssim100\,$-$\,500\,\kms
\citep[\eg][]{Libeskind06a,Campanelli07a}.
Several sources have been proposed as recoiling BH candidates \citep[][]{Komossa08a, Civano10a, Steinhardt12a}.
However, \citet{Decarli14a} have recently refuted
such scenario for one of the candidates --- SDSS J0927+2943 --- by finding that the redshift of its
BLR lines is indeed consistent with its CO systemic redshift.
This is in contrast with RXJ1131, where our CO observations confirm
the redshifted BLR lines compared to the CO systemic redshift.
Since this scenario requires a coalesced BH, it would imply that RXJ1131
is a product of a previous merger, which is not implausible and
might also explain the highly spinning BH in RX1131 \citep[$a$\ssim0.9;][]{Reis14a}.

Alternative scenarios
e.g., outflow/inflow of gas in the BLR, viewing angle towards the accretion disk, and
obscuration in the clumpy accretion disk
are more commonly invoked to explain velocity offsets between BLR and systemic redshift.
Since the BLR lines of RXJ1131 show positive velocity offsets with respect to its systemic redshift, it may imply that
the observed
BLR line emission is dominated by the gas that is flowing into the central BH, or by the receding component of
the accretion disk, owing to the viewing angle or the obscuration in the accretion disk.
\citet{Sluse07a} report a covering factor of 20\% for the accretion disk in RXJ1131
based on its broad \mgii absorption line at $z$\eq0.654.
Additionally, the centroids of the BLR lines in RXJ1131 may be biased towards longer wavelengths due to
microlensing \citep[\eg][]{Sluse07a, Sluse12a}, which may have
magnified the redshifted component of the compact BLR more strongly than its blueshifted component.

\subsection{The $M_{\rm BH}$$-$\mbulge Relation}
We find a $M_{\rm BH}$$/$$M_{\rm bulge}$ ratio of $>$0.27\petm{0.11}{0.08}\%
using the black hole mass of $M_{\rm BH}$\ssim8\E{7}\,\Msun \citep{Sluse12a}
and the stellar mass derived in \Sec{SED} as an upper limit to the bulge mass.
This ratio is consistent with those of other intermediate-$z$ radio-loud AGNs \citep{McLure06a}
but is higher than those of nearby AGNs \citep{HR04a}.
Our results therefore support
the emerging picture that quasars
grow faster and/or earlier than their host galaxies at higher redshifts \citep[e.g., ][]{Walter04a, Peng06a, McLure06a,Riechers08a}.
The elevated $M_{\rm BH}$$/$$M_{\rm bulge}$ ratio of RXJ1131 compared to local AGNs
suggests that the bulk of the black hole mass of RXJ1131 is largely in place while its stellar bulge is still assembling.

\section{Summary and Conclusions} \label{sec:sum}
We present PdBI \bco and CARMA \cco observations towards the
quadruply-imaged quasar RXJ1131 at $z_{\rm CO}$\ssim$0.654$, making this the first
resolved CO study at intermediate redshift.
Using the CO line intensities, we find a brightness temperature ratio of $r_{32}$\eq$0.78\,\pm\,0.37$
between the \bco and \cco lines,
consistent with thermalized excitation but also with the lower excitation seen in normal star-forming disks.
We also detect marginally resolved
2\,mm continuum emission underlying the \bco line
and resolved radio continuum emission at 5\,GHz in archival VLA data
in both the foreground lensing galaxy and RXJ1131.

Based on our lens modeling analysis of different \bco velocity channels,
we find a secondary CO-emitting source near RXJ1131 whose spatial position
is consistent with those of an optically faint companion reported in previous optical studies
\citepalias{Claeskens06a,Brewer08a}.
The magnification factor inferred for the CO emission in RXJ1131 is found to
vary from $\mu_{\rm L}$\ssim3 to $\sim$9 across channels. This is indicative of an extended molecular gas
distribution in the host galaxy of RXJ1131, where the different kinematic components
of the gas are magnified inhomogeneously, similar to what was found for the $z$$>$4 quasar
PSS J2322$+$1944 \citep{Riechers08a}.
Upon correcting for lensing magnification and subtracting a contribution from the companion,
we find an intrinsically symmetric double-horned \bco line profile for RXJ1131.
This together with a symmetric source-plane velocity gradient argues for a rotating disk in RXJ1131, in good agreement with previous findings \citepalias{Claeskens06a}. Physical quantities derived for RXJ1131 and the companion throughout this paper are summarized in \Tab{prop}.

Based on the lensing-corrected \bco line intensities,
we find an intrinsic gas mass of $M_{\rm gas}$\,=\,$($1.38\pmm0.33$)$\,\E{10} \Msun
for RXJ1131
and $($1.92\pmm0.09$)$\,\E{9} \Msun for the companion,
corresponding to a gas mass ratio of $\sim$7:1.
Using the source-plane \bco size of $R$\ssim6\,kpc, we find a dynamical mass of $M_{\rm dyn}$\ssim8\E{10}\,\Msun for RXJ1131.
The dynamical gas mass fraction of $f_{\rm gas}$\eq$M_{\rm gas}$/$M_{\rm dyn}$\ssim18\% and baryonic gas mass fraction of
$f_\textrm{mol}$\eq$M_{\rm gas}$/($M_{\rm gas} + M_*$)\ssim34\% are consistent with the trend of decreasing molecular gas
content since $z$\ssim2 \citep[e.g.,][;\citetalias{Combes13a}]{Lagos11a,Tacconi13a}
which has been suggested as the cause for the decline in sSFR and cosmic \SF history towards $z$\ssim0 \citep[e.g.,][]{Tacconi13a,CW13,Genzel15a}.
The CO-based dynamical mass ratio of $\sim$24:1
between RXJ1131 and the companion, and a gas mass ratio of $\sim$7:1
suggest that the system is a gas-rich, ``wet'' minor merger.

Fitting dust SED models to the IR-to-mm photometry, we derive
a lensing-corrected dust mass of $M_{\rm dust}$\ssim3\E{8}\,\Msun,
an infrared luminosity of \LIR$\sim$\,1.5\E{12}\,$(5.5/\mu_{\rm L})$\,\Lsun,
and a far-IR luminosity that corresponds to a SFR$_{\rm FIR}$\ssim120\,\sfrU.
These physical properties suggest that the merger system is dusty in nature with on-going \SF activity occuring
at a rate comparable to local ULIRGs/mergers and high-$z$ massive disk galaxies \citep{dacunha10a, Daddi10a}.
We also derive a stellar mass of $M_{*}$\ssim3\E{10}\,\Msun by fitting SED models to the
rest-frame UV-to-mm photometry, which have been corrected for their respective magnification factors before performing the fit to account for differential lensing effect.

The source-plane distribution of the gas and stellar populations of different ages
indicates that the CO gas is of similar spatial extent as the old and long-lasting stellar populations,
whereas regions of recent \SF may be embedded within the molecular gas reservoir as a result of
gas accumulation driven by interactions with the companion.
Based on dynamical mass constraints, we cannot rule out the possibility that the
compact \SF in the host galaxy
may be heavily dust-obscured.
Hence, the true extent of recent \SF may be as extended as the molecular gas
reservoir.

While properties such as CO linewidth, SFR, and gas mass found in RXJ1131
are consistent with those of local ULIRGs and high-$z$ starburst galaxies,
its SFE is comparable to those of nearby and high-$z$ disk galaxies rather than
\SB systems. This is in good agreement with its disk-like kinematic signatures and its extended molecular gas distribution.
We find a specific \SF rate (sSFR\ssim4 Gyr\pmOne) that is $\lesssim$1.5$\sigma$ higher than those of ``main sequence'' galaxies.
The slight elevation in sSFR over the main sequence suggests that
the on-going star formation activity in RXJ1131 could be enhanced by interactions with the companion.
Recent simulations have illustrated that the disk component of a gas-rich
progenitor galaxy with low SFE can be
retained upon merging since the efficiency at removing angular momentum of the gas via
gravitational torques provided by stellar components is reduced in such a system \citep{Springel05a, Robertson06a, Hopkins09a}.
As such, the extended gas disk of RXJ1131 together with its low SFE may indicate
that the \SF in RXJ1131 could form a
larger stellar bulge in the remnant disk galaxy upon coalescing.
This picture is in agreement with the one based on the \mbox{$M_{\rm BH}$$-$$M_{\rm bulge}$} relation,
where we find an elevated \mbox{$M_{\rm BH}$$/$$M_{\rm bulge}$} ratio of $>$0.27\petm{0.11}{0.08}\% for
RXJ1131 compared to the local value.
This suggests that the stellar bulge of RXJ1131 is still assembling in order to evolve onto the local relation.

We find that the redshift inferred from the NLR lines reported in previous studies are consistent with the  systemic redshift as measured
from the CO line,
but that the BLR lines are redshifted by $\sim$715\,\kms.
We raise several plausible scenarios that may explain the observed velocity offset, \eg
outflow/inflow of gas in the BLR, kinematics of the accretion disk, geometric effects,
microlensing, and a recoiling black hole from merger event.
The latter scenario might also explain the high black hole spin parameter of
$a$\eq0.87\petm{0.08}{0.15} reported by \citet[][]{Reis14a}, but
further evidence is needed to confirm or rule out this scenario.

Theoretical studies have suggested that negative feedback from an AGN may remove a
large fraction of the molecular gas from its host galaxy, thereby quenching its star formation \mbox{\citep[e.g., ][]{DiMatteo05a}}.
In this study, we find that the star formation efficiency and specific SFR of RXJ1131 are comparable to those of
$z$\ssim1$-$1.5 disk galaxies, which are not known to host quasars, and that its
molecular gas mass fraction is consistent with the observed cosmic decline for star-forming
galaxies since $z$\ssim2$-$3.
Hence, we find no evidence of negative AGN feedback on the cold molecular gas fraction
and on the star formation activity in RXJ1131.
Future observations at higher resolution will allow us to better constrain the molecular gas kinematics and dynamics of
RXJ1131 to investigate any potential interplay with the quasar on smaller scales.
More broadly, systematic studies of the correlations between the molecular gas fraction, stellar mass, and AGN luminosity
at different redshifts
will enable us to better understand the relative importance of AGN feedback and of the
evolution in the molecular gas mass fraction on the decline of star formation history and black hole accretion history.

\begin{acknowledgments}
We thank the referee for providing detailed and constructive comments that helped to improve the clarity of this manuscript.
DR and RP acknowledge support from the National Science Foundation
under grant number AST-1614213 to Cornell University. RP acknowledges
support through award SOSPA3-008 from the NRAO. DR acknowledges the
hospitality at the Aspen Center for Physics and the Kavli Institute
for Theoretical Physics during part of the writing of this manuscript.
This work is based on observations carried out under project number S14BX
with the IRAM NOEMA Interferometer. IRAM is supported by INSU/CNRS (France), MPG (Germany) and IGN (Spain).
Support for CARMA construction was derived from the Gordon and Betty Moore
Foundation, the Kenneth T. and Eileen L. Norris Foundation, the James S.
McDonnell Foundation, the Associates of the California Institute of
Technology, the University of Chicago, the states of Illinois, California, and
Maryland, and the National Science Foundation. Ongoing CARMA development and
operations are supported by the National Science Foundation under a
cooperative agreement and by the CARMA consortium universities.
The National Radio Astronomy Observatory is a facility of the National Science
Foundation operated under cooperative agreement by Associated
Universities, Inc.
This research made use of data obtained with {\it Herschel}, an ESA space
observatory with science instruments provided by European-led Principal
Investigator consortia and with important participation from NASA.
This research has made use of NASA's Astrophysics Data System Bibliographic
Services.
This work is based in part on observations
made with the NASA/ESA Hubble Space Telescope, and obtained from the Hubble
Legacy Archive, which is a collaboration between the Space Telescope Science
Institute (STScI/NASA), the Space Telescope European Coordinating Facility
(ST-ECF/ESA) and the Canadian Astronomy Data Centre (CADC/NRC/CSA).
This work is based
in part on observations made with the \spitzer,
which is operated by the Jet Propulsion Laboratory, California Institute of
Technology under a contract with NASA.
This publication made use of data products from the Wide-field Infrared
Survey Explorer, which is a joint project of the University of California, Los
Angeles, and the Jet Propulsion Laboratory/California Institute of Technology,
funded by the National Aeronautics and Space Administration.
This publication made use of data products from the Two Micron All Sky
Survey, which is a joint project of the University of Massachusetts and the
Infrared Processing and Analysis Center/California Institute of Technology,
funded by the National Aeronautics and Space Administration and the National
Science Foundation.
This research made use of the NASA/IPAC Extragalactic Database (NED) which
is operated by the Jet Propulsion Laboratory, California Institute of
Technology, under contract with the National Aeronautics and Space
Administration.
This research made use of Astropy, a community-developed core Python package for Astronomy \citep{astropy}.
This research made use of APLpy, an open-source plotting package for Python hosted at \url{http://aplpy.github.com}.

Facilities: IRAM PdBI, CARMA, VLA, Herschel(SPIRE), WISE, IRAS, 2MASS, Spitzer(IRAC, MIPS), HST(ACS, NICMOS)
\end{acknowledgments}

\bibliographystyle{yahapj}
\bibliography{RXJ}
\end{document}